\documentclass[twocolumn]{aastex631}

\usepackage{enumerate,multirow}

\def\eg{{\it e.g.}}
\def\etal{{\it et al.}}
\def\etc{{\it etc.}}
\def\ie{{\it i.e.}}

\def\Msun{M$_\odot$}

\def\pmb#1{\setbox0=\hbox{$#1$}%
  \kern-0.25em\copy0\kern-\wd0
  \kern.05em\copy0\kern-\wd0
  \kern-0.025em\raise.0433em\box0}
\def\spmb#1{\setbox1=\hbox{${\scriptstyle #1}$}%
  \kern-0.25em\copy1\kern-\wd1
  \kern.05em\copy1\kern-\wd1
  \kern-0.025em\raise.0433em\box1}

\usepackage{graphicx}	

\shorttitle{Global stability of disks}
\shortauthors{Sellwood \& Carlberg}

\begin{document}

\title{The stability of some galaxy disks is still perplexing}

\author{J. A. Sellwood}
\affiliation{Steward Observatory, University of Arizona, 933 N Cherry Ave,
  Tucson AZ 85722, USA}
\email{sellwood@arizona.edu}

\author{R. G. Carlberg}
\affiliation{Department of Astronomy and Astrophysics, University of Toronto,
  ON M5S 3H4, Canada}
\email{raymond.carlberg@utoronto.ca}

\begin{abstract}
The problem of how some disk galaxies avoid forming bars remains
unsolved.  Many galaxy models having reasonable properties continue to
manifest vigorous instabilities that rapidly form strong bars and no
widely-accepted idea has yet been advanced to account for how some
disk galaxies manage to avoid this instability.  It is encouraging
that not all galaxies formed in recent cosmological simulations
possess bars, but the {\em dynamical explanation} for this result is
unclear.  The unstable mode that creates a bar is understood as a
standing wave in a cavity that reflects off the disk center and the
corotation radius, with amplification at corotation.  Here we use
simulations to address one further idea that may perhaps inhibit the
feedback loop and therefore contribute to stability, which is to make
the disk center dynamically hot and/or to taper away mass from the
inner disk, which could be masked by a bulge.  Unfortunately, we find
that neither strategy makes much difference to the global stability of
the disk in the models we have tried.  While deep density cutouts do
indeed prevent feedback through the center, they still reflect
incoming waves and thereby provoke a slightly different instability
that again leads to a strong bar.
\end{abstract}


\section{Introduction}
\label{sec.intro}
Bars are a common feature of disk galaxies: visual classifications
\citep[\eg][]{Will13} suggest 25\%-30\% of disk galaxies are barred,
consistent with older estimates of the strongly barred fraction
\citep[reviewed by][]{SW93}.  Remarkably, the estimated barred
fraction rises to near 70\% in galaxy images taken in the near-IR
\citep{Buta15, E18}.  However, these surveys all agree that a
substantial minority of well-observed disk galaxies lack a bar, which
presents the problem we address here.  For example, \cite{SSL} were
unable to account for the absence of a bar in M33, a particularly well
studied galaxy having a gently rising inner rotation curve, a heavy
disk, and two prominent spiral arms but no large bar.  These
properties of M33 are by no means unique; the SPARC sample
\citep{Lelli16} includes several unbarred galaxies such as NGC3877,
Holmberg~IV, and UGC11557 that also have both heavy disks and gently
rising rotation curves.

\subsection{Disk instabiilities}
The bar instability of massive disks has been known for many years
\citep{Hohl71, OP73}.  These early simulations were buttressed
by \citet{Kaln78}, who presented a global mode analysis of the
full-mass isochrone disk, and his predicted mode was confirmed in
simulations by \citet{ES95}.  \citet{SA86} used the dominant mode of
the Kuzmin-Toomre disk as a test case for their mode-fitting
procedure, and \citet{Toom81} provided examples of the modes of the
Gaussian disk.  In all these cases, the massive disks of the
unperturbed models had gently rising inner rotation curves.

\citet[][see also \citealt{BT08}]{Toom81} convincingly accounted for
the instability as a cavity mode, with feedback through the center and
amplification at corotation, which we describe more fully in
\S\ref{sec.mechsm}.  He also predicted that cutting the feedback loop,
\eg\ by inserting a dense bulge, which would cause waves to be damped
at an inner Lindblad resonance (ILR), could stabilize the entire disk,
as \citet{Zang76} had apparently found for linear instabilities having
$m\geq2$ in the Mestel disk.

\citet{ELN} studied how bar-formation in an exponential disk having
constant $Q$ was affected by changes to the rotation curve.  They
reported that bar-stability was dependent on the amplitude of the
rotation curve, reaching the much-cited conclusion that the halo in
bar-stable models should be ``the dominant contributor to the total
mass.''  Embedding the disk in a dense halo suppresses the bar
instability because $m=2$ waves are no longer amplified.  However,
this is not a satisfactory explanation for the absence of bars in
galaxies because the disk should then manifest multi-arm spiral
patterns; two-armed spirals, which are the most common patterns in
galaxies \citep{Davis12, Hart16, YH18}, would be suppressed by the
dominant halo for the same reason. 

\citet{ELN} also reported that disk stability was independent of the
halo core radius, although that second finding was privately
challenged by Toomre because it was at variance with his suggestion
that a dense bulge should cause most reasonable patterns to be damped
at an ILR.  \citet{Sell89} indeed found that the small-$N$ simulations
by \citet{ELN} were affected by non-linear amplified shot noise that
overwhelmed the ILR and, in more careful experiments, he confirmed
Toomre's contention that a dense bulge-like mass can stabilize the
disk.  However, the insertion of an ILR is not a panacea for the bar
instability, partly because few galaxies can be sufficiently smooth
and quiescent over a long period of time for all swing-amplified
disturbances to be damped at an ILR and partly because \citet{Sell12}
later reported that even initially very smooth simulations of the
Mestel disk suffered from secular growth of non-axisymmetric stuctures
due to impedance changes in the bulk of the disk, caused by weak
resonant scattering, which eventually led to a strong bar.

Furthermore, \citet{Athan02, Atha08}, \citet{SN13}, and \citet{BS16}
found that the bar instability is yet more vigorous in simulations
that employ a halo composed of mobile particles, rather than a rigid
mass distribution.  This is because the global mode in the disk is
able to elicit a supporting response from the halo that varies in the
expected manner with the anisotropy of the halo velocity distribution
\citep{Sell15}.

Disks also support other types of mode for which the mechanism is not
a standing wave.  The most notable are edge \citep{Toom81, PL89} and
groove \citep{SK91} modes that are driven from the corotation
resonance, but neither creates a bar.  Bars can also be formed by modes
related to the radial orbit instability \citep[\eg][]{LB79, PP94} that
may operate in globally stable disks, but these ideas do not help to
account for the absence of bars in some disks.

\citet{BLLT} presented a global stability analysis of a large family
of disk-halo models, finding that those having cool, low-mass disks
and dense bulges supported slowly growing spiral modes.  Those authors
interpreted the instabilities as cavity modes also, with the more mild
being of the type proposed by \citet{Mark77}. These WASER modes invoke
travelling waves {\em refracting} off a ``$Q$-barrier'' from the
short- to the long-wave branches of the WKB dispersion relation
\citep[see][]{BT08, SM22}.  Though both trailing waves, the short- and
long-waves propagate radially in opposite directions, allowing a
feed-back loop that was closed by mild amplification at corotation.
As far as we are aware, the only direct test of one of their cases was
presented by \citet{Sell11} who reported a long-lived wave having a
constant pattern speed in a simulation in which disturbance forces
were restricted to $m=2$.  However, \citet{Sell11} also reported that
more vigorous instabilities rapidly emerged in the same model when
disturbance forces from higher sectoral harmonics were included.

\subsection{Cosmological simulations}
Cosmological simulations with hydrodynamics are developing apace and
create objects that bear some resemblance to galaxies \citep[for a
  review, see][]{CV23}.  \citet{Algo17} and others have examined the
frequency and properties of bars in the simulated galaxies, finding
the bar fraction to be somewhat lower than that observed.  However,
this could simply be due to inadequate resolution, as \citet{Zhou20}
find a higher bar frequency in the Illustris TNG100 models, especially
among the higher mass galaxies \citep{Zhao20}.

Furthermore, \citet{Algo17}, \citet{Rosh21}, and others have noted
that bars in these simulations generally have too low a pattern speed,
in the sense that the corotation radius is much larger than the bar
semi-major axis, which differs from the properties of observed bars in
the nearby universe \citep[][see also \citealt{Butt23}]{Ague15}.  Once
again, \citet{Fran22} find the discrepancies of bar properties between
simulations and observations are lessened, but not as yet eliminated,
as the resolution of the simulations is improved.

Slow bars in the simulations are probably a consequence of dynamical
friction from too high a density of dark matter near the centers of
galaxies \citep[][and much subsequent work]{Wein85, DS00}.
\citet{Mara20} highlighted halo domination as a shortcoming of the
simulations on other grounds.  Furthermore, \citet{Nava18} studied a
multi-arm spiral disk from their simulation, which as already noted,
is a symptom of an overly dominant halo; the relatively weak heating
and radial migration in the model they studied is likely also a
consequence of the mild, multi-arm spirals in that sub-maximal disk.

This very incomplete summary of the relevant literature indicates that
the frequency of barred ``galaxies'' in cosmological simulations is
perhaps lower than among galaxies in the local universe, and the bars
have properties that differ from those of observed galaxies.  However,
trends with improving resolution suggest some of these differences may
ultimately go away.  But the reasons that a substantial fraction are
unbarred include (1) stabilization by overly dense halos
\citep[\eg][]{Redd22}, (2) that infalling sub-halos may destroy bars,
though that would both heat and thicken the disk of the host galaxy.
As far as we are aware, this is a topic that has yet to be thoroughly
addressed \citep[but see \eg][]{Ghos21}.  A final possibility (3) is
that there is some as yet unknown stabilizing factor at work in both
the simulations and in the real universe that prevents some galaxies
from forming bars.

But in relation to the question we address in this paper, we do not
find that cosmological simulations have yielded any clear
understanding of why a galaxy has, or lacks, a bar.  Indeed
\citet{Zhou20} concede in their summary that the morphologies of
``individual galaxies are subject to the combined effects of
environment and internal baryonic physics and are often not
predictable.''  It is important to keep improving the resolution, and
perhaps also tweaking the feedback recipe \etc, in order to reproduce
the observed frequency and properties of bars, but understanding from
such complicated simulations the {\em dynamical reason} for the
eventual match of the models with the observed facts will be extremely
challenging.  We therefore pursue a parallel investigation using
idealized models in which we have some hope of developing deeper
insight into this complicated question of disk dynamics.

\subsection{This paper}
We note that the extensive set of models presented by \citet{BLLT}
explored regimes that have not otherwise been carefully examined.
While employing the same type of galaxy model as did \citet{ELN}, they
considered disks that were dynamically hot in the center and/or in
which the inner disk surface density had been tapered away, implying a
rigid bulge-like component to maintain the adopted rotation curve.
They reported that these properties had profound effects on the shapes
and growth rates of the dominant linear instabilities.

Our purpose here is to test whether bar-formation can be averted in a
moderately heavy stellar disk,\footnote{Loosely, more than a
half-maximum disk say, \ie\ one that contributes, at its peak central
attraction, a fraction that is not much less than that of the spherical
matter at the same radius.} either by increasing the velocity spread
of the stars in the inner disk and/or by reducing the surface density
of the inner disk.  We therefore present a stability study of models
that resemble, but do not exactly match, a small subset of those
explored by \citet{BLLT}.  We find fairly vigorous bi-symmetric global
modes in most cases with pattern speeds high enough to avoid ILRs.
The linear bar-mode can be tightly wrapped in the inner part, while
feedback clearly includes a reflection from trailing to leading.
Reflection off the center is inhibited in strongly cutout disks, but a
new type of instability replaces the classic bar-mode.  In both cases,
the dominant mode again leads to a large bar.

\section{Mass models}
\label{sec.models}
Following \citet{BLLT}, we consider a family of idealized disk-halo
galaxy models in which we determine the dominant instability.  In our
case, we use 2D quiet-start \citep{Sell83} $N$-body simulations to
follow the evolution of the initial equilibrium model, and fit the
dominant mode to the simulation data using the technique described by
\citet{SA86}.

\subsection{Baseline model}
\label{sec.basic}
Our baseline model is of the type originally proposed by \citet{FE80}.
The surface density of a flat, axisymmetric exponential disk has the
radial profile
\begin{equation}
  \Sigma(R) = \Sigma_0 e^{-R/R_d} \quad\hbox{with}\quad \Sigma_0 = {M_d
    \over 2\pi R_d^2},
\label{eq.expdisk}
\end{equation}
where $R_d$ is the disk scale length and $M_d$ is the nominal mass of
the infinite disk.  We limit its radial extent using a cubic function
to taper the surface density from $\Sigma(5R_d)$ to zero at $R=6R_d$.

The rotation curve is that of a cored isothermal sphere
\begin{equation}
  V(r) = V_0 \left[ { r^2 \over r^2 + r_c^2} \right]^{1/2},
\label{eq.Vcirc}
\end{equation}
with $r_c$ being the core radius, although we will be interested
exclusively in quantities in the disk plane where $r=R$.  The implied
halo density is whatever is required, when combined with the disk
attraction, to achieve this rotation curve in the disk plane.  We
relate the rotation curve to the disk properties by setting $V_0 = 0.9
(GM_d/R_d)^{1/2}$, and generally choose $r_c = 0.5R_d$.  As the
maximum circular speed arising from a razor-thin exponential disk is
$0.622(GM_d/R_d)^{1/2}$, the disk, though quite heavy, has less than
the required mass to account for the central attraction at any radius.

Since we are here interested in bisymmetric linear instabilities of
the disk only, we generally do not compute the axisymmetric central
attraction of the disk.  We compute only the $m=2$ disturbance forces
from the disk particles, and then add the central acceleration
$-V^2(r)/r$ to every particle at every step.  Note that this strategy
implies that we represent the halo component as a rigid mass
distribution.

We set the radial velocity dispersion of the disk particles using the
\citet{Toom64} criterion
\begin{equation}
  \sigma_R(R) = Q(R) \sigma_{R,\rm min}, \quad\hbox{where}\quad
  \sigma_{R,\rm min} = {3.36 G\Sigma \over \kappa},
\label{eq.dispsn}
\end{equation}
and $\kappa$ is the local epicyclic frequency \citep{BT08}.  Note that
we generally use a radially dependent $Q$ function, although in all our
models we set $Q_{\rm OD} = 1.2$ in the outer disk.

We adopt the distribution function for the disk component using the
form proposed by \citet{Shu69}
\begin{equation}
  f(E,L_z) = \cases{{\cal F}(L_z) e^{-{\cal E}/\sigma_R^2(R_g)} & $0<{\cal
    E} \leq -E_c(L_z)$, \cr 0 & $L_z <0$. \cr}
\label{eq.ShuDF}
\end{equation}
Here ${\cal E}$ is the excess energy of a particle above $E_c$, which
is that of a circular orbit at the guiding center radius $R_g(L_z)$.
Although this DF assumes no retrograde stars, we later reverse the
angular momentum of some low-$L_z$ particles in order to smooth the
discontinuity in $f(E,L_z)$ at $L_z=0$, which does not affect the
equilibrium.  The function (\ref{eq.ShuDF}) clearly assumes a Gaussian
velocity distribution at all radii, which \citet{Shu69} argues is the
appropriate form for a (partially) relaxed disk.  We select particles
from this DF using the method described in the appendix of
\citet{DS00}.

The function ${\cal F}(L_z)$ has to be determined numerically and the
procedure we adopt is described in the on-line manual \citep{Sell14}.
As there are many possible functions ${\cal F}$ that fit the adopted
disk surface density, we impose two extra requirements.  Not only are
rapid fluctuations of ${\cal F}$ with $L_z$ physically unreasonable,
but we have also found that even mild ``ripples'' in the function
${\cal F}(L_z)$ can seed additional disk instabilities related to
groove modes \citep{SK91}.  We therefore penalize the fit to the
surface density also to minimize
\begin{equation}
  T = \sum_{L_z}\left[ {d^2{\cal F} \over dL_z^2} \right]^2.
\label{eq.smDF}
\end{equation}
Note that the numerical search for the optimum ${\cal F}$ seeks a
balance between fitting the disk surface density while also minimizing
$T$, and finding the optimum balance is something of an art.  It is
also required that ${\cal F}(L_z) \geq 0$ for all $L_z$, although we
find that this requirement is generally satisfied for a smooth ${\cal
  F}$ without imposing an additional constraint.

In order to create a quiet start, we place three copies of each
particle almost regularly around a half-circle.  By restricting
disturbance forces to $m=2$ only, these three particles mimic an
initially smooth, circular wire of uniform mass per unit length that
osillates radially at the epicyclic frequency and distorts in response
to particle dynamics in the global gravitational potential.

Here, and throughout the paper, we use units such that $G=M_d=R_d=1$.
Our unit of time is therefore $\tau_{\rm dyn}=(R_d^3/GM_d)^{1/2}$.
For those who prefer physical units, a possible scaling is to set
$R_d=2\;$kpc, and $\tau_{\rm dyn}=10\;$Myr, which implies $V_0 \simeq
176\;$km~s$^{-1}$ and $M_d \simeq 1.78\times 10^{10}\;$\Msun.

\subsection{Variants of the baseline model}
\label{sec.variants}
Following \citet{BLLT}, we consider models with prescribed
$Q$-profiles and disks having central mass cutouts, and in two cases
we also increase the halo core radius $r_c$.

\begin{figure}
\begin{center}
\includegraphics[width=.8\hsize,angle=0]{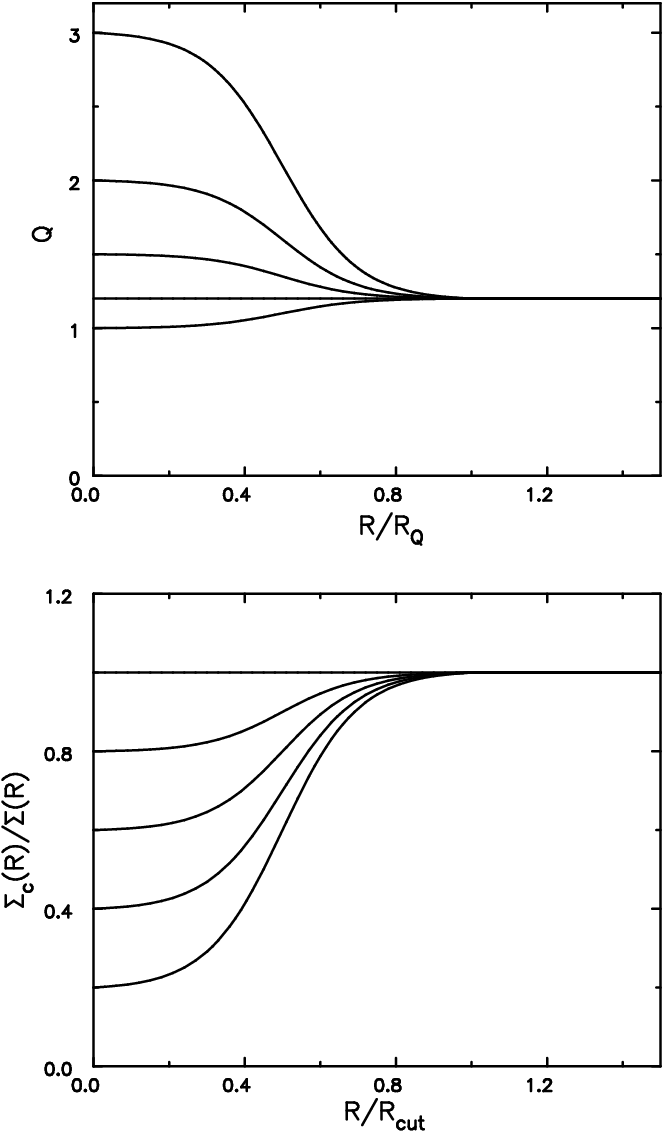}
\end{center}
\caption{The upper panel shows the function $Q(R)$ (eq.\ref{eq.Qprof})
  and the lower panel the ratio $\Sigma_c(R)/\Sigma(R)$
  (eq.\ref{eq.Sigtap}).  The lines in both panels use most adopted
  values of $Q_{\rm jump}$ and $D$.}
\label{fig.tapers}
\end{figure}

The $Q$-profile has the functional form
\begin{equation}
  Q(R) = \cases{ Q_{\rm OD} + Q_{\rm jump} {\cal T}(x) & $R<R_Q$ \cr
                 Q_{\rm OD} & otherwise, \cr}
\label{eq.Qprof}
\end{equation}
where $R_Q$ is the radial extent of the higher $Q$ values, $Q_{\rm
  jump}$ is the difference between $Q(0)$ and $Q_{\rm OD}$, the
argument $x = 1 - 2R/R_Q$ varies from $1 \geq x \geq -1$ and the
Fermi-like taper function
\begin{equation}
{\cal T}(x) = {1\over2}\left[ 1 + {(e^{5x}+1)^{-1} + 0.5 \over (e^5 + 1)^{-1} - 0.5}\right],
\label{eq.Fermi}
\end{equation}
rises smoothly from zero at $x=-1$, to unity at $x=+1$.

The central surface density is cut away in the following manner
\begin{equation}
  \Sigma_c(R) = \Sigma(R) \cases{ D + {\cal T}(x)(1-D)] & $R<R_{\rm
      cut}$ \cr 1 & otherwise, \cr}
\label{eq.Sigtap}
\end{equation}
where the central depth $D = \Sigma_c(0)/\Sigma_0$, and in this case
$x = 2R/R_{\rm cut} - 1$.  Since the rotation curve is unaffected,
deeper cutouts imply the galaxy model has a greater bulge mass.  Note
that $\sigma_R$ will also be reduced as more mass is cut away (see
eq.~\ref{eq.dispsn}), provided that the prescribed function $Q(R)$ is
unchanged.

The functions eqs.~(\ref{eq.Qprof}) and (\ref{eq.Sigtap}), illustrated
in Figure~\ref{fig.tapers}, introduce four new parameters: $R_Q$ and
$R_{\rm cut}$ which are the radial extents of the enhanced $Q$ values
and of the surface density cutout, and $Q_{\rm jump}$ and $D$.  In
principle, $R_Q$ and $R_{\rm cut}$ could differ, but we have found it
convenient to keep $R_Q = R_{\rm cut}$, which reduces the number of
extra parameters by one.

\subsection{Simplifying assumptions}
\label{sec.assumps}
Our simulations embody several approximations to reduce the
computational cost.  As this is a preliminary search for stabilizing
influences, any models that may turn out to be promising would need to
be resimulated without these approximations.  Specifically:
\begin{itemize}
\item Particles are restricted to motion in a plane as a first
  approximation.  Note large-scale non-axisymmetric modes are only
  slightly weaker in a moderately thickened disk than in a 2D disk,
  and gravity softening provides some allowance for disk thickness.
\item The restriction of disturbance forces to a single sectoral
  harmonic does not matter, because each sectoral hamonic behaves
  independently until the amplitude becomes large, and here we wish to
  measure the mode frequency in the linear regime.  We confirm in
  \S\ref{sec.cutout} that including extra harmonics does not alter the
  linear mode and causes differences at large amplitude only.
\item We employ a rigid halo since we are searching for stabilizing
  influences.  If a disk is unstable with a rigid halo, then it has
  no chance of being stable in a live one, but any apparently stable
  cases could be re-simulated later using a live halo.
\item We have also neglected the role of gas. Recall \citet{SSL}
  reported that the global stability of the disk of M33 was very
  little affected when the observed gas component was treated either
  as collisionless particles or as gas in a number of different
  ways. Note that gas comprises $\ga 30$\% of the total disk mass in
  that galaxy, though the atomic component is more spread out than
  are the stars.
\end{itemize}
We critically re-examine our adopted model and simplifiying
assumptions in \S\ref{sec.discuss}.

\begin{table}
\caption{Default numerical parameters} 
\label{tab.DBHpars}
\begin{tabular}{@{}ll}
Grid points in $(r, \phi)$ & 170 $\times$ 256 \\
Grid scaling & $R_d = 10$ grid units \\
Active sectoral harmonic & $m=2$ \\
Plummer softening length & $\epsilon = R_d/20$ \\
Number of particles & $18 \times 10^6$ \\
Largest time-step & $0.2\tau_{\rm dyn}$ \\
Radial time step zones & 4 \\
\end{tabular}
\end{table}

\begin{table*}
\begin{center}
\begin{tabular}{@{}lclccccc}
& Name & $r_c/R_d$ & $R_Q=R_{\rm cut}$ & $Q(0)$ & $D = {\Sigma_c(0) \over \Sigma_0}$ & $m\Omega_p$ & $\beta$  \\ 
\hline
\multirow{3}{*}{Increasing $r_c$}
& baseline & 0.5 &  N/A & 1.2 & 1.0 & $0.802\pm0.001$ & $0.031\pm0.000$ \\
& Rc0.75 & 0.75 &  N/A & 1.2 &  1.0 & $0.629\pm0.000$ & $0.032\pm0.001$ \\
& Rc1.0 & 1.0 &  N/A &  1.2 &   1.0 & $0.587\pm0.000$ & $0.035\pm0.001$ \\
\hline
\multirow{3}{*}{Hot disk center}
& HC1.0 & 0.5 &  1.5 &  1.0 &   1.0 & $0.867\pm0.006$ & $0.033\pm0.001$ \\
& HC1.5 & 0.5 &  1.5 &  1.5 &   1.0 & $0.712\pm0.001$ & $0.030\pm0.001$ \\
& HC2.0 & 0.5 &  1.5 &  2.0 &   1.0 & $0.644\pm0.001$ & $0.023\pm0.001$ \\
\hline
\multirow{3}{*}{Disk cutout 1}
& 1DC0.8 & 0.5 &  1.5 &  1.2 &  0.8 & $0.779\pm0.002$ & $0.035\pm0.001$ \\
& 1DC0.5 & 0.5 &  1.5 &  1.2 &  0.5 & $0.838\pm0.001$ & $0.039\pm0.001$ \\
& 1DC0.2 & 0.5 &  1.5 &  1.2 &  0.2 & $0.826\pm0.001$ & $0.061\pm0.001$ \\
\hline
\multirow{4}{*}{Disk cutout 2}
& 2DC0.8 & 0.5 &  1.5 &  2.0 &  0.8 & $0.672\pm0.000$ & $0.029\pm0.000$ \\
& 2DC0.6 & 0.5 &  1.5 &  2.0 &  0.6 & $0.783\pm0.000$ & $0.037\pm0.001$ \\
& 2DC0.4 & 0.5 &  1.5 &  2.0 &  0.4 & $0.781\pm0.000$ & $0.050\pm0.001$ \\
& 2DC0.2 & 0.5 &  1.5 &  2.0 &  0.2 & $0.776\pm0.000$ & $0.061\pm0.001$ \\
\hline
\multirow{3}{*}{Increasing $R_Q$}
& RQ1.0 & 0.5 &  1.0 &  3.0 &   0.2 & $0.860\pm0.001$ & $0.046\pm0.001$ \\
& RQ1.5 & 0.5 &  1.5 &  3.0 &   0.2 & $0.738\pm0.000$ & $0.051\pm0.000$ \\
& RQ2.0 & 0.5 &  2.0 &  3.0 &   0.2 & $0.637\pm0.000$ & $0.043\pm0.000$ \\
\hline
\multirow{3}{*}{Miscellaneous}
& MQ1.5 & 0.5 &  1.5 &  1.5 &   0.2 & $0.798\pm0.000$ & $0.057\pm0.001$ \\
& MR2.5 & 0.5 &  2.5 &  2.0 &   0.4 & $0.587\pm0.000$ & $0.025\pm0.000$ \\
& MR3.5 & 0.5 &  3.5 &  2.0 &   0.4 & \multicolumn{2}{c}{no instability at $m=2$} \\ 
\hline

\end{tabular}
\end{center}
\caption{List of simulations.  The horizontal lines break the runs
  into sequences in which a single parameter is varied.  N/A values in
  col 3 are because both $Q_{\rm jump}=0$ and $D=1$.  The small quoted
  uncertainties give the spreads in fits to a single simulation, but
  more realistically are $\pm\sim10\%$ in both the real and imaginary
  parts -- see \S\ref{sec.fits}.}
\label{tab.runs}
\end{table*}

\subsection{{\sc Galaxy} code}
Previous work \citep{Sell83, SA86, ES95, SE01} has established that
simulations with a 2D polar grid can reproduce the predicted linear
instabilities of a variety of mass models.  We therefore use this
method to determine the dominant modes of these new models.  The
particles move over a 2D polar mesh and their mutual gravitational
attractions are calculated at grid points and interpolated to the
position of each particle.  A full description of our numerical
procedures is given in the on-line manual \citep{Sell14} and the
source code is available for download.  Table~\ref{tab.DBHpars} gives
the values of the numerical parameters adopted for most simulations
presented in this paper.  Fourier analysis of the mass distribution on
each grid ring, which separates the solution for the field into
different sectoral hamonics, makes it easy to restrict the disturbance
forces to those arising from $m=2$ distortions of the particle
distribution.  We report checks in which we vary these parameters in
\S\ref{sec.innref} below, and a test with a 2D Cartesian grid in
\S\ref{sec.Cartes}.

\begin{figure}
\begin{center}
\includegraphics[width=.9\hsize,angle=0]{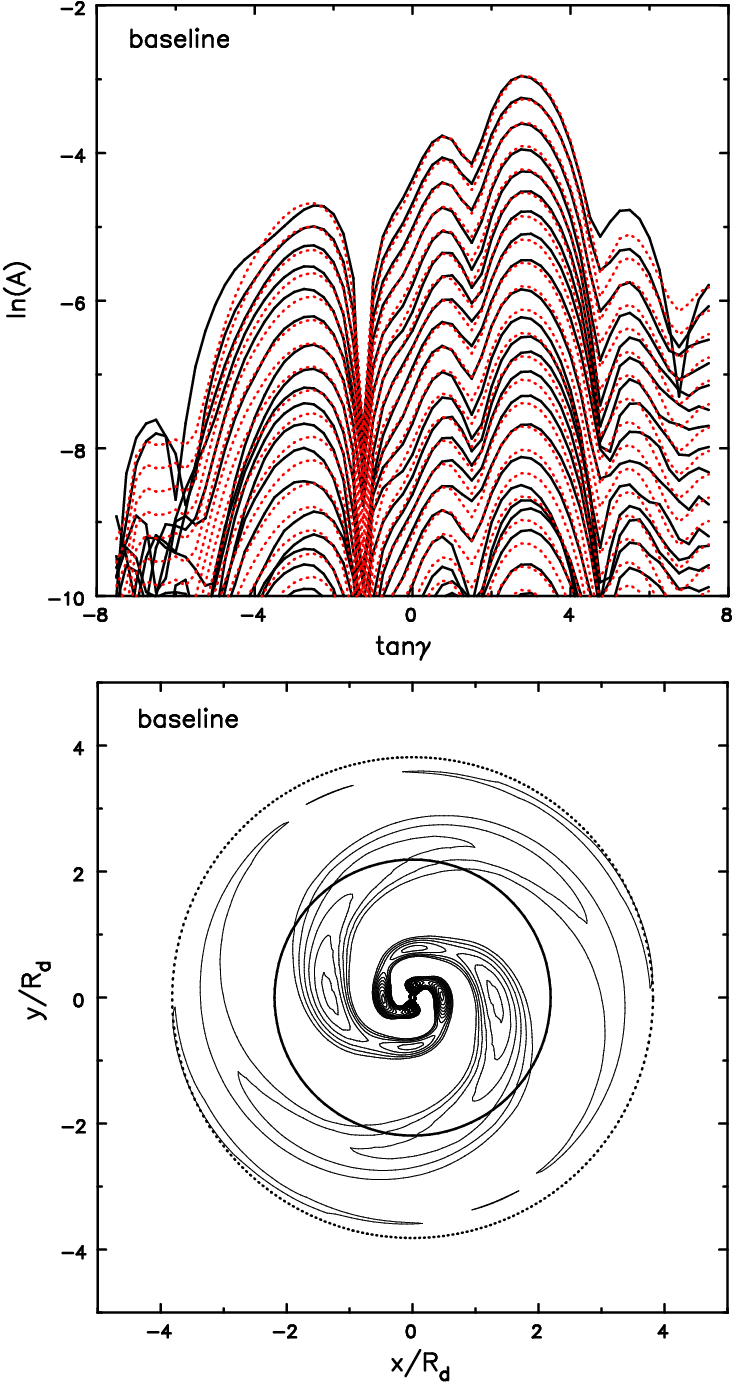}
\end{center}
\caption{The black lines in the upper panel present the amplitude of
  the logarithmic spiral transform of the particle positions
  (eq.~\ref{eq.logspi}) at intervals of $10\tau_{\rm dyn}$ in the
  baseline model.  The red, dotted curves show the fit of a single
  gowing mode to the data at the same times.  The contours in the
  lower panel are of the best fit mode taken from a fit over the same
  time interval as above, but to the density transforms on the grid
  rings.  The full-drawn and dotted circles mark respectively the
  radii of the CR and the OLR -- the pattern speed of this mode is
  high enough to be well clear of possible ILRs.}
\label{fig.example}
\end{figure}

As usual, we measure non-axisymmetric distortions of the distribution
of the $N$ particles using an expansion in logarithmic spirals:
\begin{equation}
A(m,\gamma,t) = {1 \over N}\sum_{j=1}^N \, \exp[im(\phi_j + \tan\gamma \ln R_j)],
\label{eq.logspi}
\end{equation}
where $(R_j,\phi_j)$ are the polar coordinates of the $j$th particle
at time $t$, $m$ is the sectoral harmonic, and $\gamma$ is the
(radially constant) angle of the spiral component to the radius
vector, which is the complement to the spiral pitch angle.

\subsection{Mode fitting}
\label{sec.fits}
Recall that a normal mode is a self-sustaining, sinusoidal disturbance
of fixed frequency and constant shape.  The perturbed surface density
of a mode in a galaxy disk is the real part of
\begin{equation}
\delta\Sigma(R,\phi,t) = A_m(R)e^{i(m\phi - \omega t)},
\label{eq.mode}
\end{equation}
where the frequency, $\omega$, is complex when the mode grows or
decays. The complex function $A_m(R)$, which is independent of time,
describes the radial variation of amplitude and phase of the mode.
Here, $\omega = m\Omega_p + i\beta$, with $\Omega_p$ being the pattern
speed and $\beta$ the growth rate.

The results in Table~\ref{tab.runs} were obtained by fitting the
function (eq.~\ref{eq.mode}) to the data from the simulation using the
least-squares procedure described by \citet{SA86}.  The input data to
the fit are either the logarithmic spiral transforms
(eq.~\ref{eq.logspi}) or the amplitude and phase of the $m=2$
component of the mass assigned to the points on each grid ring.  A
quiet start reduces the seed amplitude of the mode to well below the
level that would be expected from particle shot noise, and is
essential to allow a long enough period of linear growth to obtain a
credible measurement of the growth rate.  In a few cases, a fit of two
superposed modes seems to be preferred, but we list only that with the
highest growth rate.

We quote uncertainties in the measured frequencies in
Table~\ref{tab.runs} that span the entire spread of values from fits
to the data of both types and over slightly different time ranges in a
single simulation.  However, they are probably severe underestimates
because different realizations of the same model turn out to have
unstable mode frequencies that differ by some $\sim 10\%$.  Choosing
different parameter values, such as the weight given to the smoothing
term $T$ in eq.~(\ref{eq.smDF}), results in a different realization
that is not exactly the same model.  Even though the fitted surface
density and $Q$-profiles of two or more realizations are barely
distinguishable, we found that minor local differences in the density
of particles as a function of $L_z$ affect the frequency of the
fundamental mode to a surprising extent.

Inner Lindblad resonances (ILRs) are avoided whenever $m\Omega_p >
(m\Omega - \kappa)_{\rm max}$, where $\Omega(R) = V(R)/R$ is the
angular frequency of circular motion and $\kappa$ is the radial
epicyclic frequency.  From eq.~(\ref{eq.Vcirc}), we find $\Omega^2 =
V_0^2/(R^2 + r_c^2)$ and $\kappa^2 = V_0^2(2R^2+4r_c^2)/(R^2 +
r_c^2)^2$ and therefore, with our adopted values: $m=2$, $V_0 = 0.9$
and $r_c = 0.5$, we find $(m\Omega - \kappa)_{\rm max} \simeq 0.382$
at $R \simeq 0.748$.

\subsection{WKB waves}
We interpret the properties of the fitted modes in terms of a local
dispersion relation for density waves in disks, which connects the
frequency of a {\em steady} wave $\omega$ to its radial wavenumber
$|k|$.  The formula is independent of the sign of $k$, which is
conventionally taken as positive for trailing waves and negative for
leading. It embodies the WKB approximation that computes the
self-gravity of the spiral as that of a plane wave in a thin sheet
\citep{BT08}.  The version given by \citet{LS66} is
\begin{equation}
\left[m(\Omega_p - \Omega)\right]^2 = \kappa^2 - 2\pi G\Sigma |k| {\cal F},
\label{eq.WKB}
\end{equation}
which states that the self-gravity term decreases the wave frequency
$\omega = m(\Omega_p - \Omega)$, here assumed to be purely real, below
the natural frequency of radial oscillation $\kappa$.  The ``reduction
factor'' ${\cal F} \leq 1$ \citep[given by][their Appendix K]{BT08}
depends upon $Q$, $k$, and $\omega$, and quantifies the extent to
which the self-gravity term is weakened by random motion.

Eq.~(\ref{eq.WKB}) has severe limitations \citep{SM22}.  It applies to
waves that are sufficiently tightly-wrapped that the spiral pitch
angle can be neglected, allowing the radial frequency $\omega$, to be
related the Doppler-shifted frequency at which stars encounter an
$m$-fold symmetric spiral $\omega = m(\Omega_p - \Omega)$
\citep{LS66}.  It also applies equally to leading and trailing waves
because it also omits any hint of swing-amplification \citep{Toom81}
near corotation.  Despite these limitations, the short wave branch
\citep{BT08} does give some qualitative indication of the behaviour
spiral waves away from corotation.

\citet{Toom69} pointed out that a spiral wave packet propagates
radially across the disk at a group velocity $v_g = \partial \omega /
\partial k$, which may be computed from eq.~(\ref{eq.WKB}).  He found
that short waves travel away from or toward the corotation resonance
(CR) when they are respectively trailing or leading and that, other
things being equal, $v_g$ should be decreased somewhat when $Q$ is
raised.  His numerical calculations showed that eq.~(\ref{eq.WKB})
breaks down over a broad region around the CR, but it makes reasonably
accurate predictions elsewhere.  The wave carries angular momentum at
the group velocity.

\begin{figure*}
\includegraphics[width=.6\hsize,angle=270]{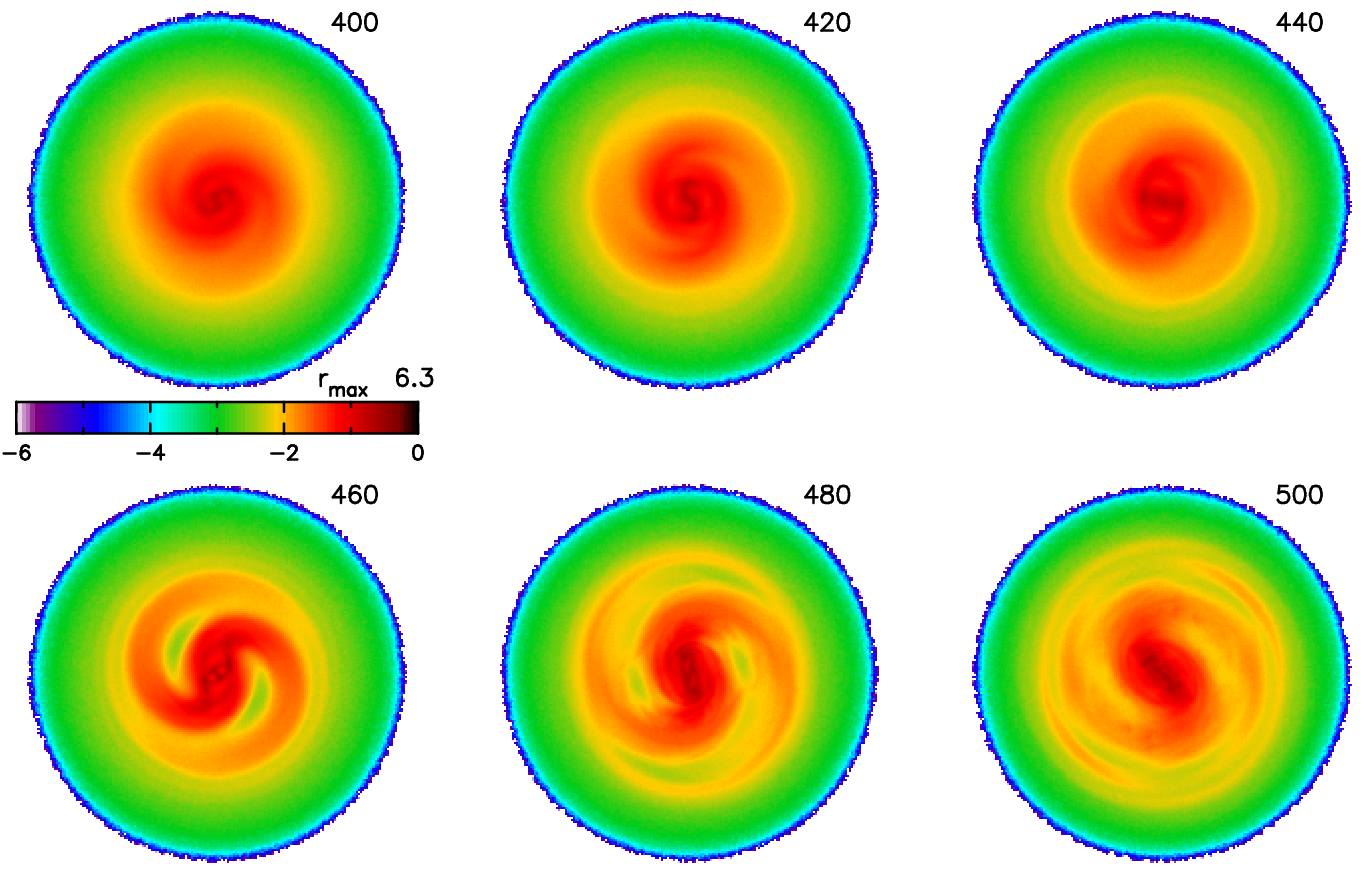}
\caption{The evolution of the baseline model from $400 \leq t \leq
  500$, which forms a strong bar.  The color scale shows the logarithm
  of the mass surface density, which initially drops to zero at
  $R=5R_d$.  Notice the inner spiral and bar features, and the tightly
  wrapped trailing spiral near the OLR in the last two panels.}
\label{fig.bar}
\end{figure*}

\section{The mode of our baseline model}
\label{sec.example}
Our baseline model is globally unstable, consistent with the earlier
study by \citet{ELN}.  The disk is massive enough and the rotation
curve rises slowly enough that a global cavity mode of the type
described by \citet{Toom81} must be expected.

\subsection{Linear mode}
Figure~\ref{fig.example} illustrates the fitting procedure for the
simulation of our baseline model.  The solid black curves in the upper
panel show the amplitude of the logarithmic spiral transforms
(eq.~\ref{eq.logspi}) of the particle positions at intervals of
$10\tau_{\rm dyn}$ during the period of linear growth --
\ie\ excluding early times that were noise-dominated, and later times
when the mode saturates.  The logarithmic amplitude scale reveals
approximate equal spacing of these curves over time, indicating steady
exponential growth.  The red, dotted curves mark the amplitude of the
fitted function (eq.~\ref{eq.mode}) at the same times, and are in
reasonable agreement with the data.  Note that the data used in the
fit are from transforms taken five times more frequently than those
illustrated.  A bias towards higher amplitude of trailing waves
($\tan\gamma>0$) over the leading components ($\tan\gamma<0$) is
evident, but the mode has significant amplitude over the range $-5 \la
\tan\gamma \la 5$.

We also fitted a mode to the density transforms on the grid rings over
the same time interval, and draw the fitted function $A_m(R)$ in the
lower panel; note that the contours are of positive overdensity only,
the underdense part is not contoured.  We have drawn a solid circle to
mark the radius of the corotation resonance (CR), and a dotted circle
at the radius of the outer Lindblad resonance (OLR); the pattern speed
of this, and all other modes in this paper (Table~\ref{tab.runs}), is
high enough that there are no ILRs.

The mode has a tightly wrapped spiral appearance in the inner parts
(lower panel of Fig.~\ref{fig.example}), that differs from the open
spiral of the domininant mode of fully self-gravitating disks that
has been reported by others.  We will show that the different
apppearance of the linear mode is largely due to the adopted rotation
curve, which raises the orbital frequencies in the inner disk above
those that arise in most self-gravitating disk models.  The mode in
Fig.~\ref{fig.example} is more open in the outer parts, however.

\begin{figure*}
\begin{center}
\includegraphics[width=.3\hsize,angle=0]{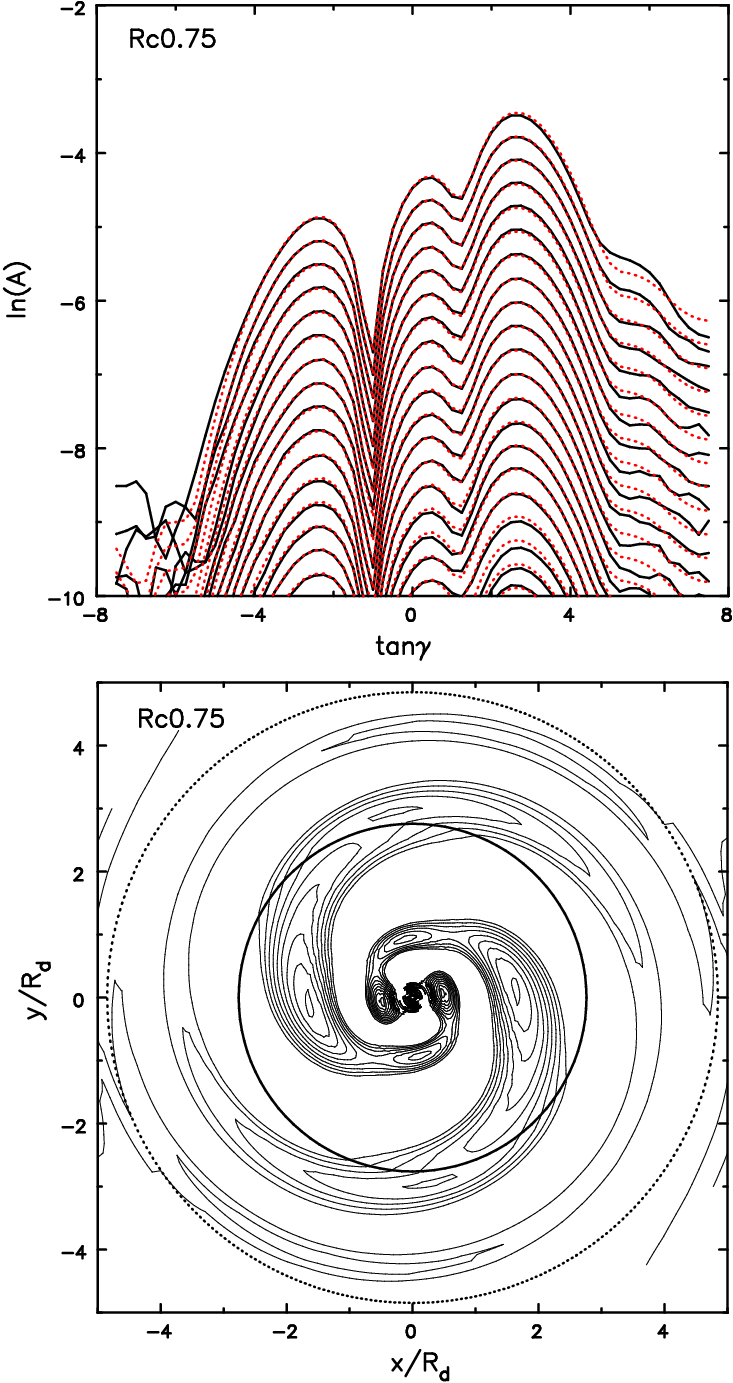}
\includegraphics[width=.3\hsize,angle=0]{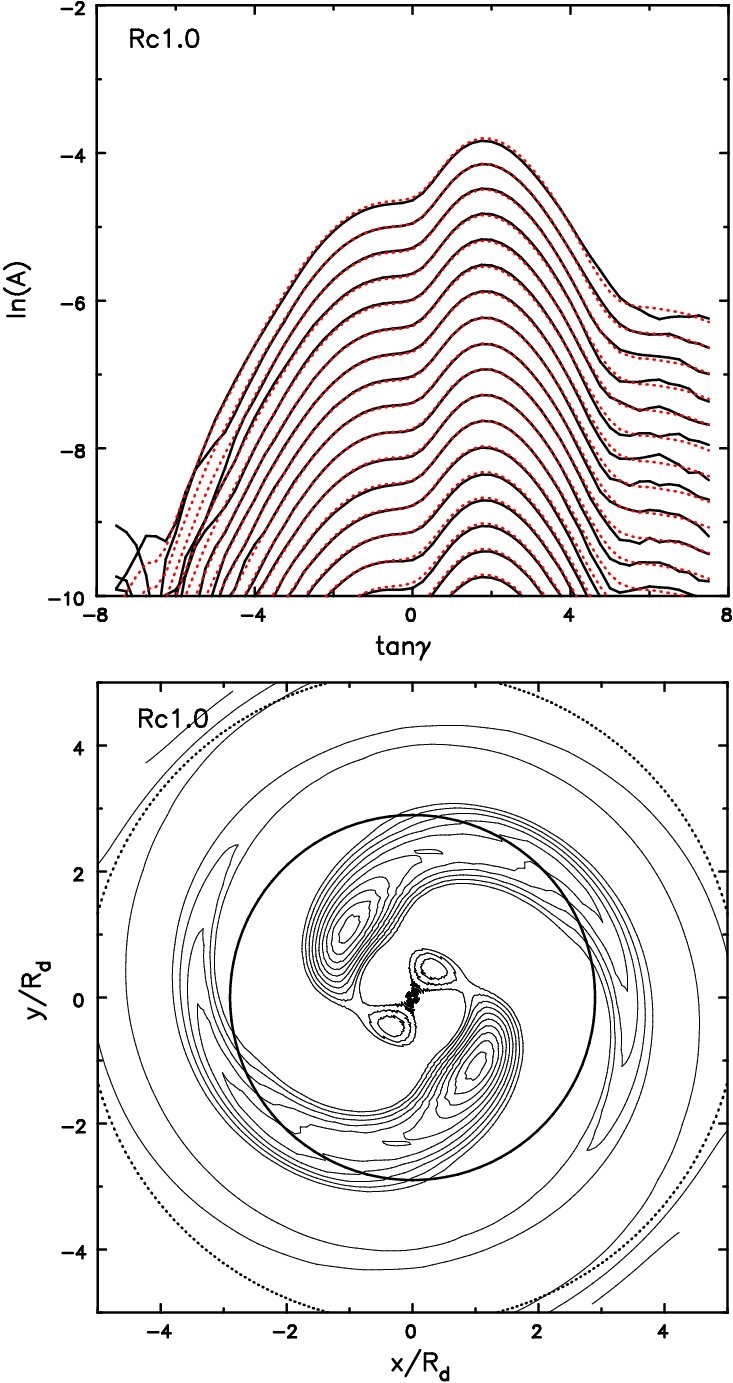}
\caption{Fits to the linear evolution of two simulations in which
  $r_c=0.75R_d$ (left) and $r_c=R_d$ (right).  As in the baseline
  model, both have flat $Q=1.2$ and no cutouts.  The two logarithmic
  spiral transforms of the mode (upper panels) differ qualitatively,
  as do the mode shapes in the inner parts (lower panels).}
\label{fig.core}
\end{center}
\end{figure*}

\newpage
\subsection{Bar mode mechanism}
\label{sec.mechsm}
As noted in the introduction, \citet{Toom81} elucidated the mechanism
of a bar mode as that of a cavity mode, or standing wave, between the
disk center and corotation.  More specifically, a trailing wavetrain
propagates inward at the group velocity and reflects off the center
into an outwardly propagating leading wavetrain, but the second
reflection off the CR causes the wavetrain to be strongly
swing-amplified into an amplified trailing wave, providing positive
feedback that leads to an unstable run away -- but only until the
failure of the small-amplitude approximation that underlies the
feedback loop.  The growing wave in the cavity between CR and the
center removes angular momentum from the inner disk \citep{LBK, BT08,
  SM22}, and since the net angular momentum of an isolated disk cannot
change, the outer disk must accept that removed from the inner disk by
the disturbance.  An exclusively trailing wave carrying positive
angular momentum propagates outward where it wraps more tightly as it
is finally absorbed by wave-particle interactions at the OLR
\citep{LBK}.

It seems plausible that this mechanism operates in the mode reported
in Fig.~\ref{fig.example}, since the disturbance reaches the very
center of the disk, where it reflects.  The bias of the leading
components of the transform in the upper panel being weaker than the
trailing components, as expected from swing amplification at the CR,
gives rise to the overall trailing appearance of the disturbance.  But
the fundamental modes of disks having gently rising rotation curves do
not manifest the multiple sub-peaks along the mode ridge line that we
find here.  These sub-peaks are regularly spaced 90$^\circ$ apart,
suggesting interference between the leading and trailing waves,
similar to that \citet{Toom81} reported for the overtone modes of the
Gaussian disk.  The mode transform in the upper panel also has
more structure than is typically seen for open bar modes, doubtless
as a consequence of interference between the leading and trailing
waves.

\begin{figure*}
\includegraphics[width=.3\hsize,angle=0]{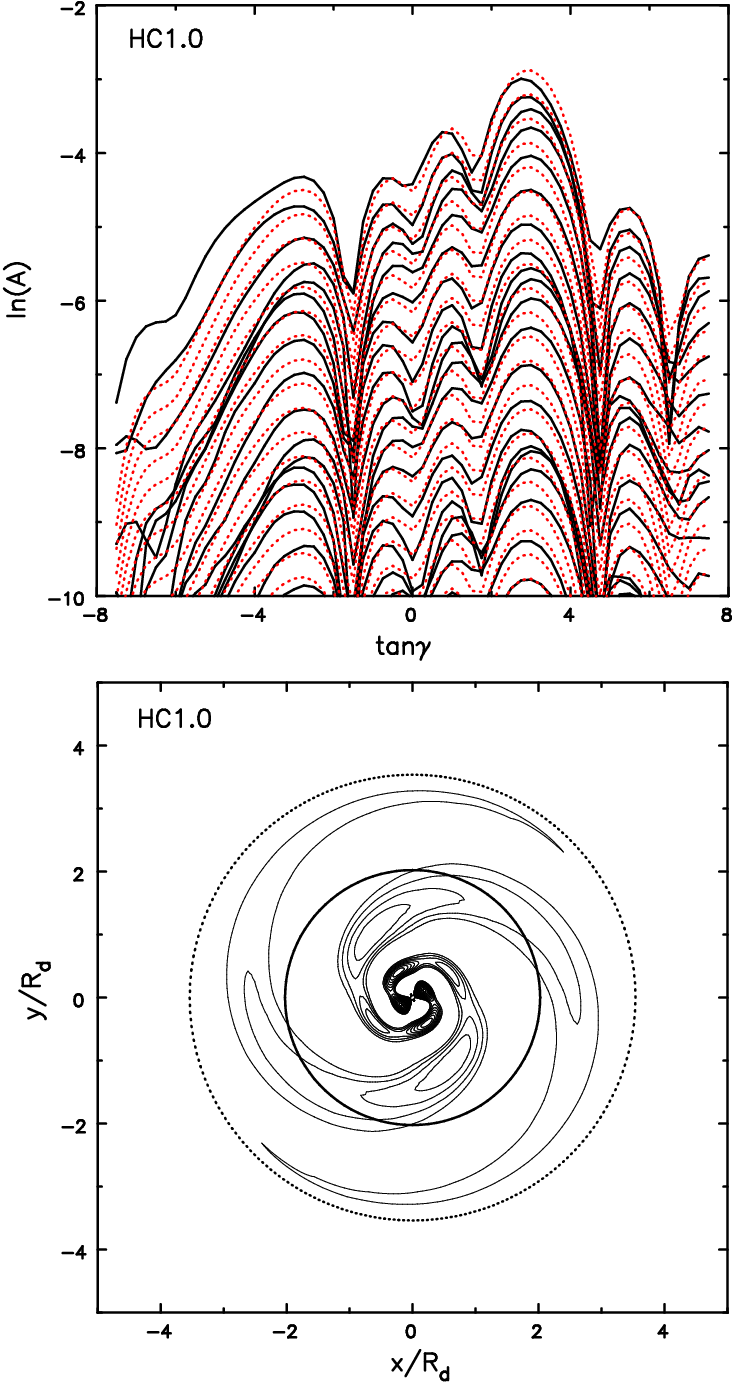}
\includegraphics[width=.3\hsize,angle=0]{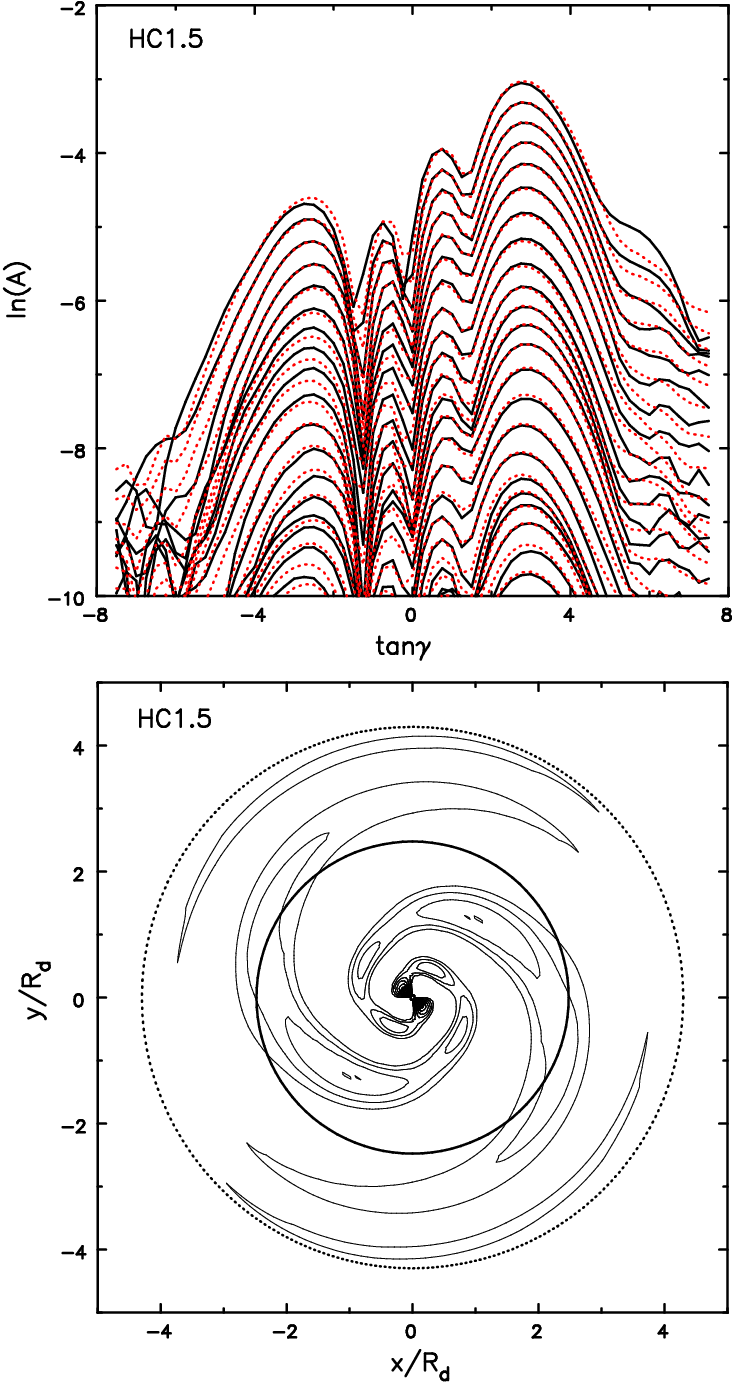}
\includegraphics[width=.3\hsize,angle=0]{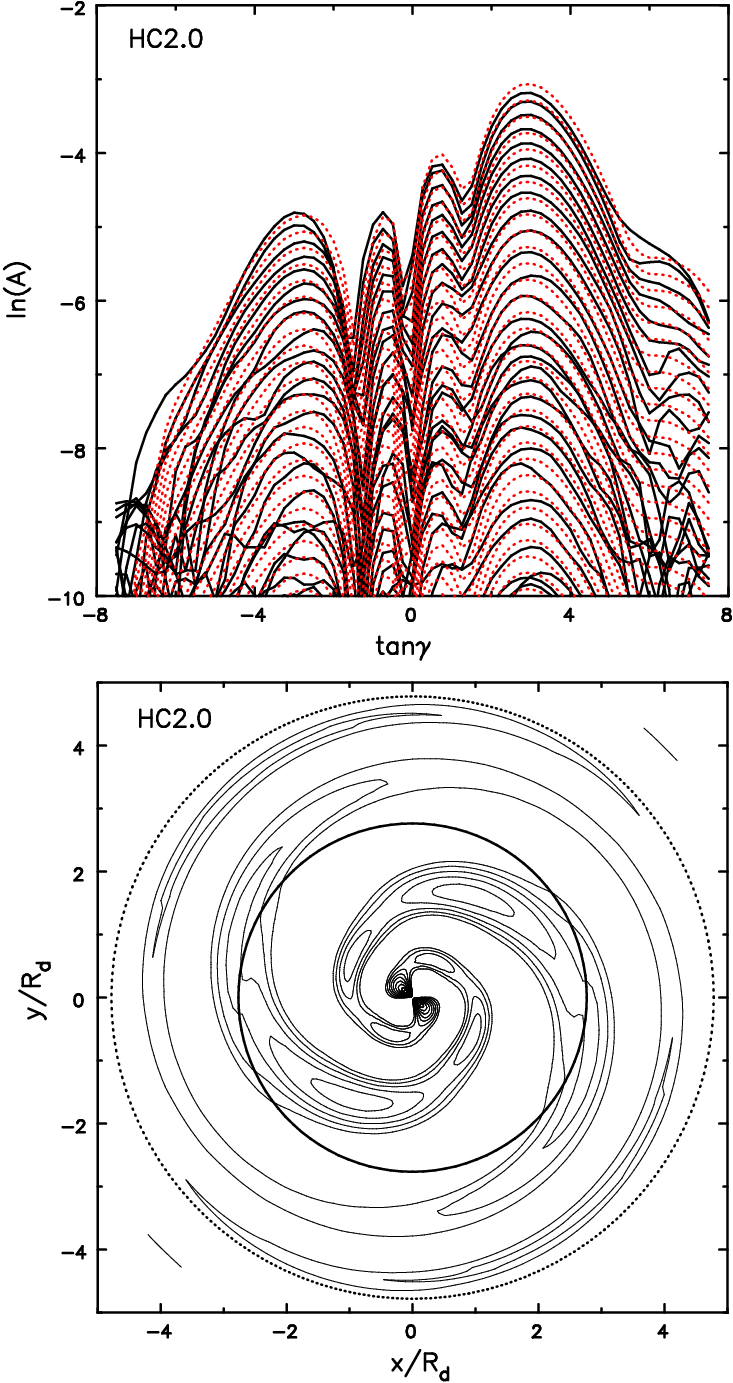}
\caption{Fits to the linear evolution of a sequence of simulations in
  which the central value of $Q(0)$ is increased; in all cases,
  $R_Q=1.5$ and there is no cutout.  There was a hint of a second mode
  in the data for the LH panel but the fit presented is for a single
  mode.  We have fitted two modes to the data in the upper middle
  panel, but present only the dominant mode in the lower middle panel.
  The increasing size of the mode shapes in the lower panels reflects
  their decreasing pattern speeds, which are given in
  Table~\ref{tab.runs}.  The growth rates of the modes in HC1.0 and
  HC1.5 are about the same as that in the baseline model, but the mode
  HC2.0 (right) grows less rapidly.}
\label{fig.risingQ}
\end{figure*}

The frequency of the mode is determined by the phase closure
constraint, which means that the complete feedback cycle must
encompass an integral number of wave periods, else the travelling
waves will self-interfere.  Normally, in simple full-mass disk models
such as the isochrone, the dominant mode is the fundamental with a
single antinode in the cavity.  Overtones having two, three, and more
antinodes were reported by \citet{Toom81} for the Gaussian disk, but
they all had lower growth rates than the fundamental.  In our case, we
see several antinodes already in this, the dominant mode of our baseline
model, presumably because the time required for the wavetrain to
travel from CR to the center and back is far longer than the
oscillation period of the wave.  But it unclear to us why this
particular mode, of all overtones that are possible in this model,
should have the highest growth rate, and therefore stand out in our
simulation.

The mechanism proposed by \citet{Toom81} suggests that the mode growth
rate is determined by the amplification factor at the CR divided by
the time required for the wavetrain to travel to the inner reflection
radius and back to the CR.  However, it would be hard to predict what
this should be because the group velocity of the wavetrain
\citep{Toom69, BT08, SM22} cannot even be estimated from WKB theory
for part of the required radial range because eq.~(\ref{eq.WKB}) has
no solutions in the ``forbidden region'' when $Q>1$.  While it is
known that growing waves do propagate in this region, we do not have a
dispersion relation for them from which we could estimate their $v_g$.

\subsection{Non-linear evolution}
Fig.~\ref{fig.bar} presents snapshots from the later part of the
baseline simulation that show the non-linear evolution, in which
perturbing forces are still restricted to $m=2$.  Forces from other
sectoral hamonics may change the appearance slightly in this
non-linear regime (see \S\ref{sec.cutout}).  It is clear that the
model forms a strong bar by $t=500$, but it is interesting that the
first feature to become visible at $t=400$ is the small inner spiral
which later settles into a short inner bar that rotates with the large
bar.

\begin{figure*}
\includegraphics[width=.3\hsize,angle=0]{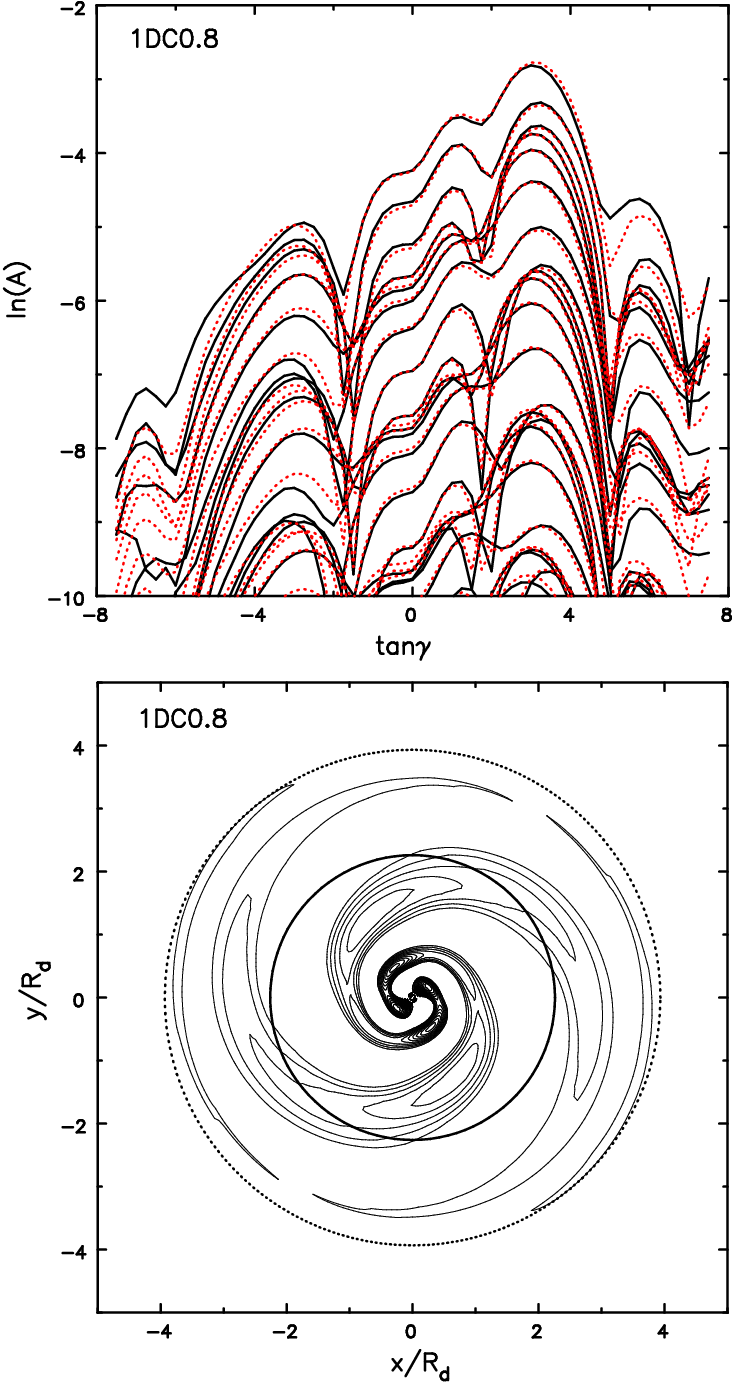}
\includegraphics[width=.3\hsize,angle=0]{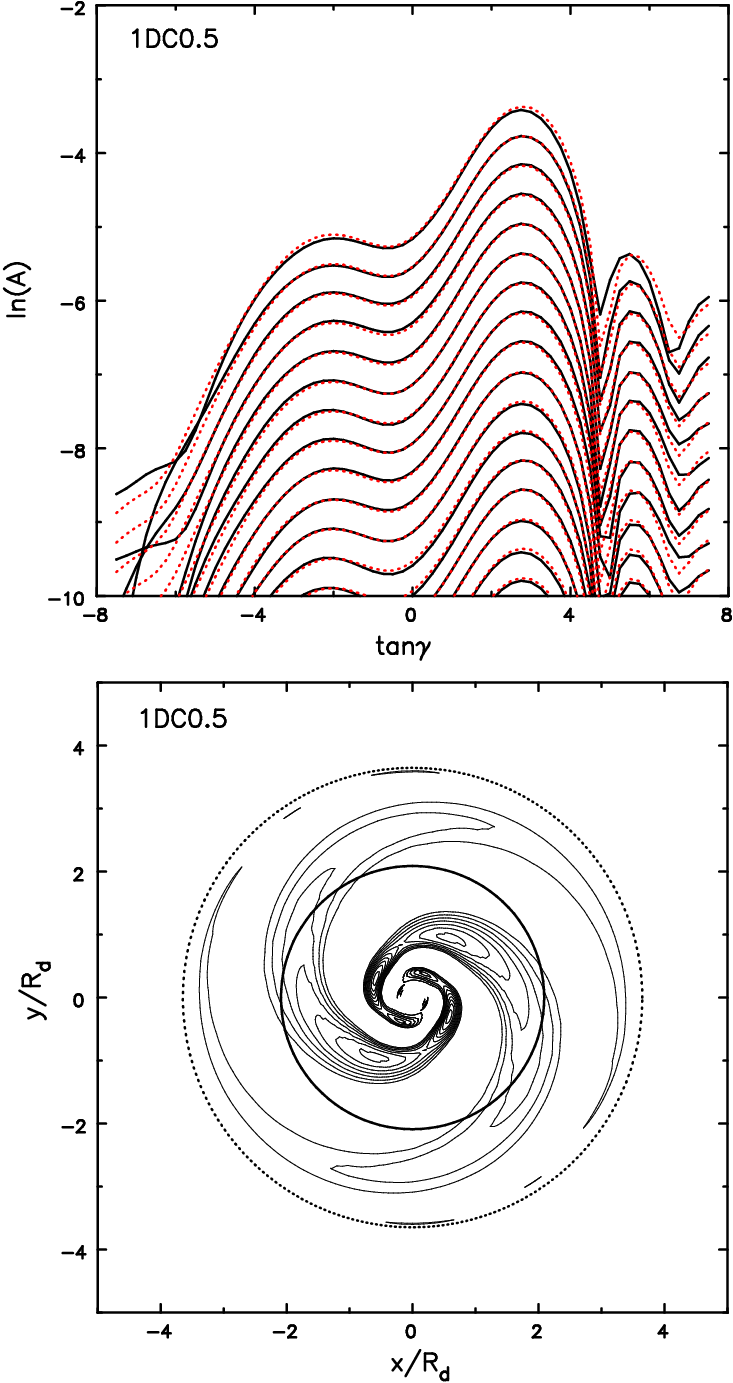}
\includegraphics[width=.3\hsize,angle=0]{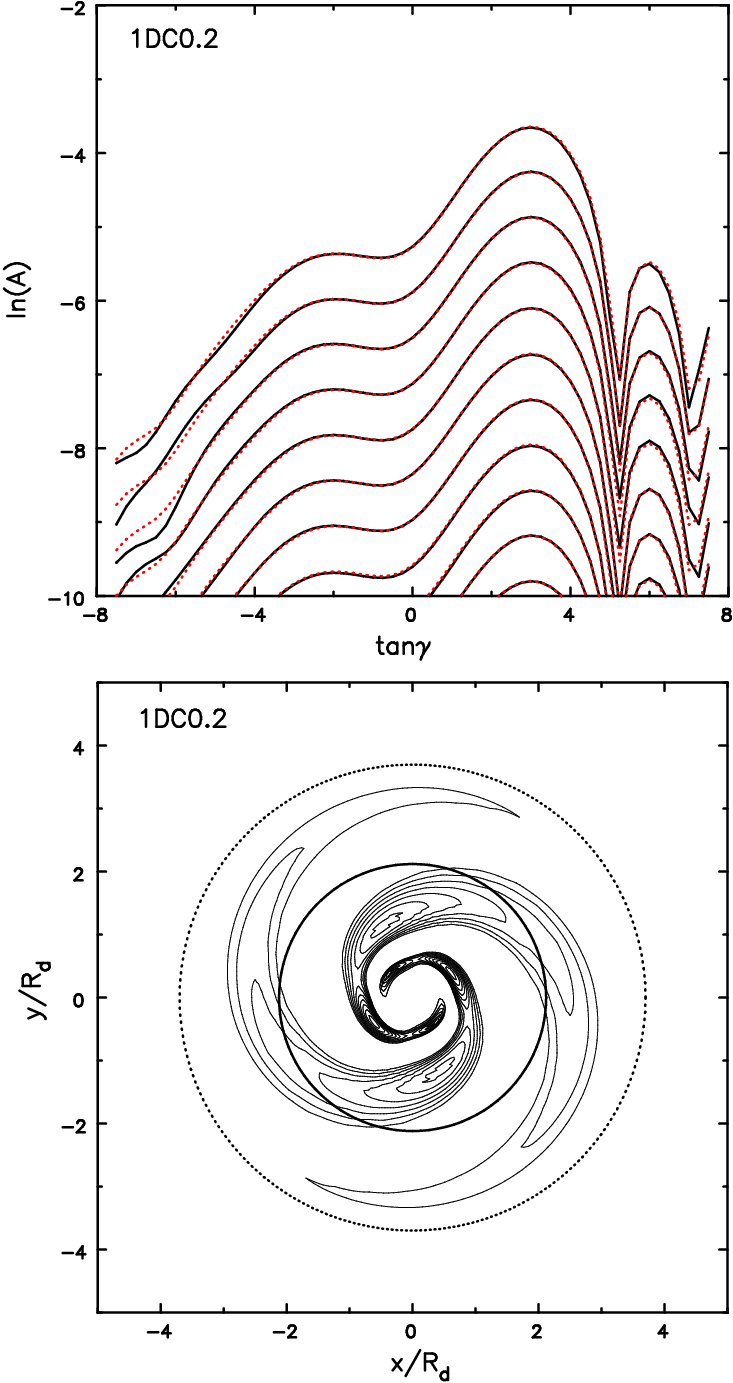}
\caption{Fits to the linear evolution of three simulations having
  increasingly deep central cutouts.  In all cases $R_{\rm cut} =
  1.5$, while $Q(R) = 1.2$ everywhere, and the baseline model
  (Fig.~\ref{fig.example}) having no cutout is the first of this
  sequence.}
\label{fig.varyD}
\end{figure*}

\subsection{A further numerical check}
\label{sec.Cartes}
It seemed possible that the choice of overtone referred to above, and
other aspects of the mode, could have been preferred by our polar
grid.  We therefore reran the same model using a 2D Cartesian grid,
with two-fold rotational symmetry imposed to suppress odd sectoral
harmonics and a strategy to kill forces that may arise from $m=4$
density variations, as described by \citet{Sell20}.  In order to
maintain the same central attraction as on the polar grid, we assigned
the mass of a smooth, tapered exponential disk to a separate copy of
the grid and then subtracted this unchanging smooth disk from the
masses of the moving particles that were assigned to the working grid
at each step, before solving for the gravitational attraction of the
residual density variations \citep{Sell20}.  Thus the accelerations
acting on the particles are those arising from the disturbance density
distribution together with the fixed attraction implied by the
rotation curve, as in our baseline simulation.  The six-fold
rotational symmetry of the quiet start, does not affect the dynamics
on the polar grid, where forces were restricted to $m=2$, but is a
nuisance on the Cartesian grid.  Therefore, from the same basic file
of particles, we placed 20 instead of three on each half-ring to make
the unperturbed mass distribution 40-fold symmetric, which suppressed
all likely non-axisymmetric forces except for those arising from
bisymmetric disturbances.

We employed a square grid with $1024$ cells on a side, set the disk
scalelength $R_d = 80$ mesh spaces and the Plummer gravity softening
length $\epsilon=R_d/20=4$ mesh spaces to be consistent with
disturbance forces on the polar grid.  Our numerical concern was
allayed when we found that the dominant disturbance in this simulation
had the same pattern speed, to within 0.2\%, growth rate to within
1\%, and mode transform that was indistinguishable from that in our
baseline run on the polar grid (Fig.~\ref{fig.example}, top panel).

\begin{figure*}
\includegraphics[width=.3\hsize,angle=0]{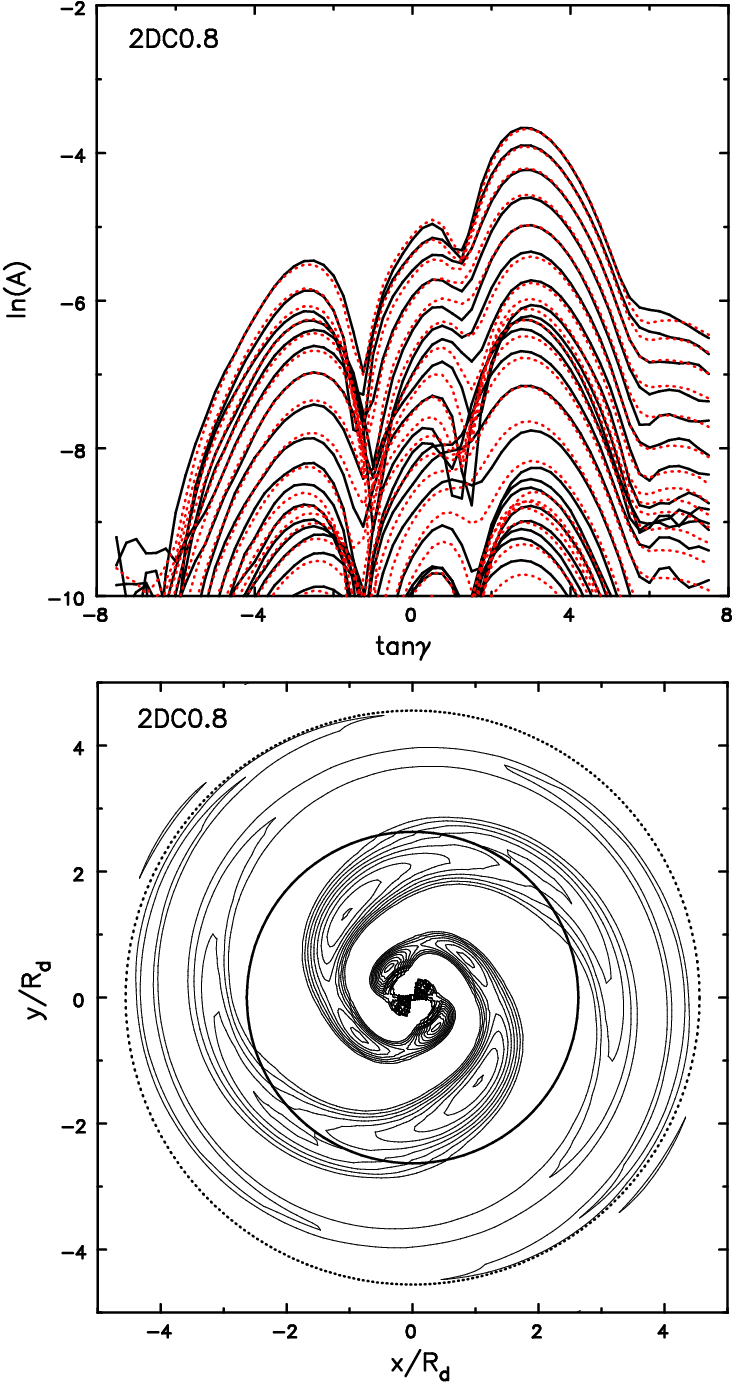}
\includegraphics[width=.3\hsize,angle=0]{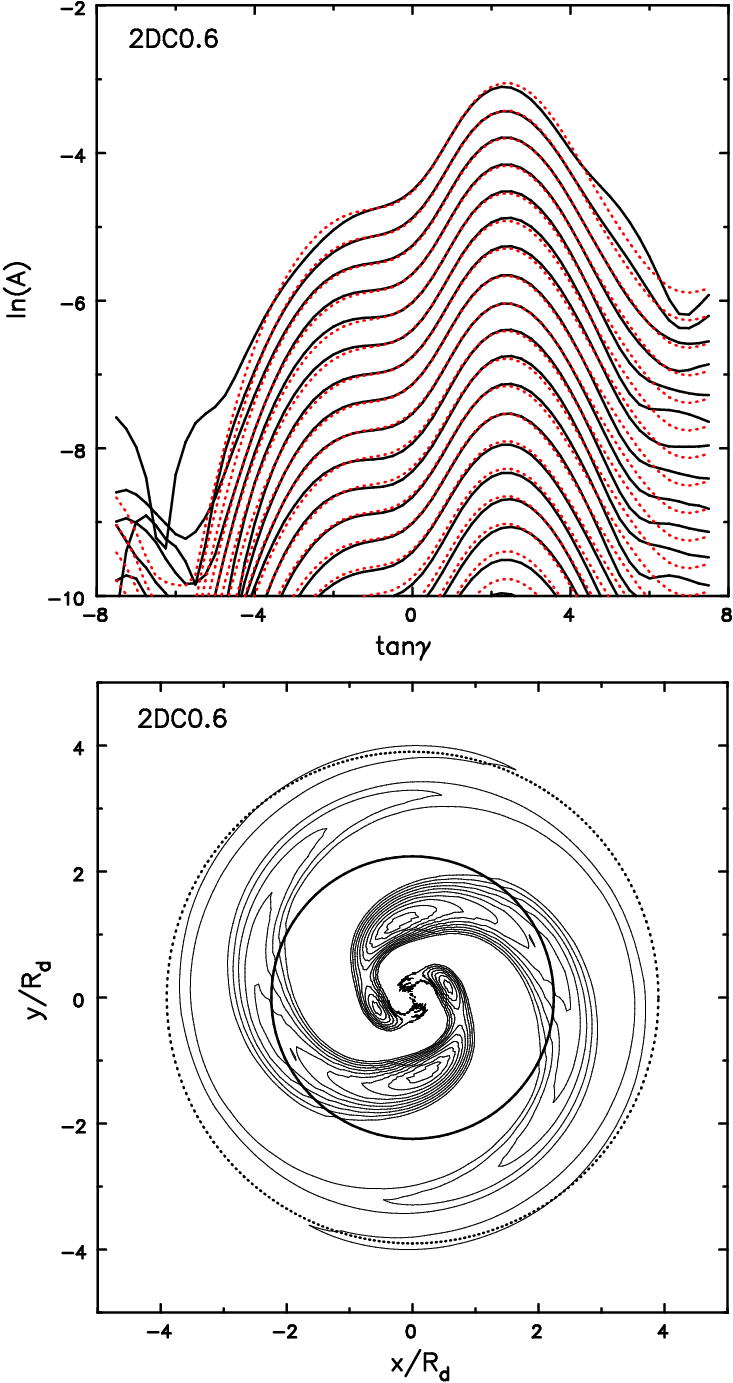}
\includegraphics[width=.3\hsize,angle=0]{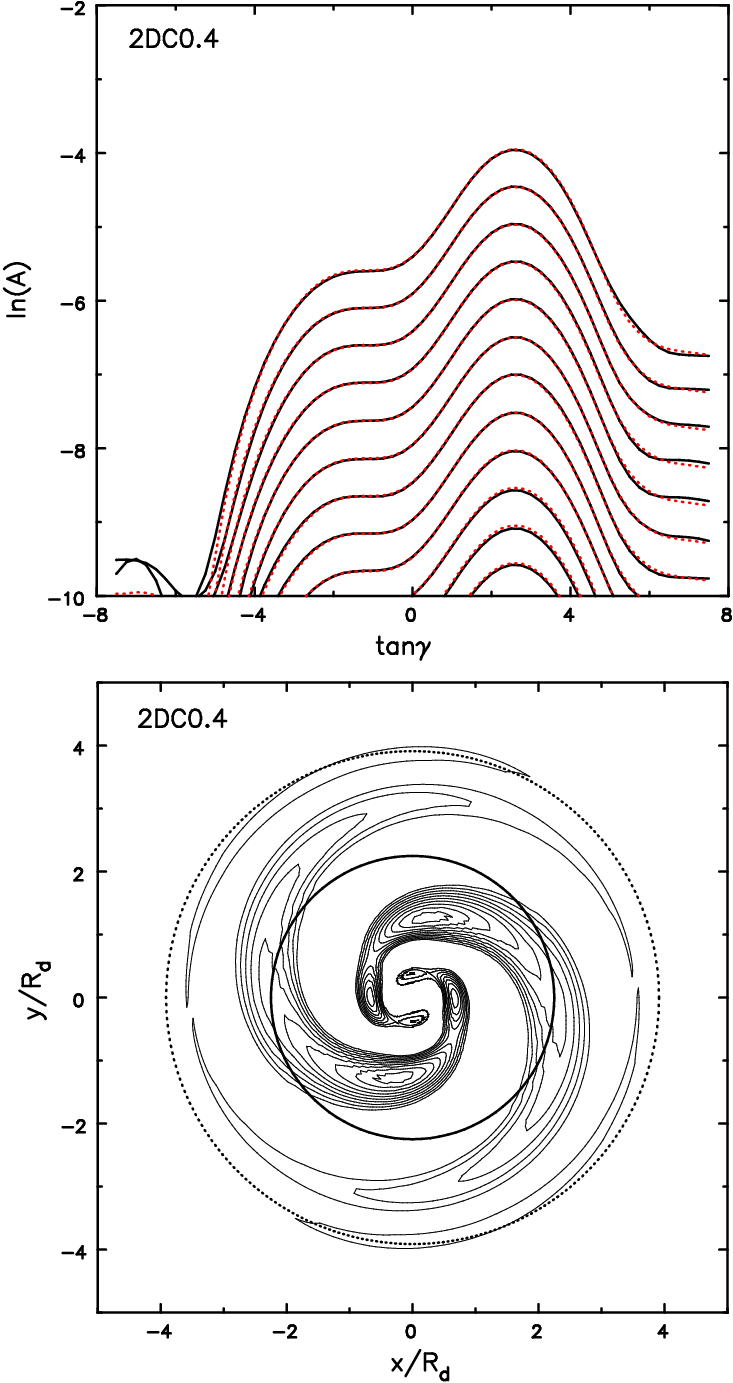}
\caption{A parallel sequence to that in Fig.~\ref{fig.varyD} but with
  the central value of $Q$ also raised to $Q(0)=2$. Note $R_{\rm cut}
  = R_Q = 1.5$ in all three cases.  The first member of this sequence,
  is HC2.0 in the right hand panel of Fig.~\ref{fig.risingQ} and the
  final member of the sequence, 2DC0.2, is presented in
  Fig.~\ref{fig.run5315}.}
\label{fig.varyD2}
\end{figure*}

\subsection{Reducing the inner slope of the rotation curve}
We can understand the tight winding of the inner part of the mode in
Fig~\ref{fig.example} through the WKB dispersion relation.
Rearranging eq.~(\ref{eq.WKB}), we find
\begin{equation}
|k| = { \kappa^2 - \omega^2 \over 2\pi G\Sigma {\cal F}}.
\label{eq.wavenumber}
\end{equation}
Note that this is not an explicit expression for $k$, because the
factor $\cal F$ in the denominator also depends on $k$, but it
strongly suggests that one way to prefer tightly-wrapped waves, or
large $|k|$, is to increase $\kappa$.  Thus in the inner disk of our
model, the numerator in eq.~(\ref{eq.wavenumber}) is increased as
$r_c$ is decreased, which is the main reason that the mode in
Fig.~\ref{fig.example} is more tightly wrapped than modes in disks
having more gently rising rotation curves.

In order to confirm that the tight winding of the inner part of the
mode was due to the smallish core radius of the global potential, we
present in Fig.~\ref{fig.core} the fitted linear modes of models
Rc0.75 (left) and Rc1.0 (right), in which we increased the core radius
from the baseline value ($r_c=0.5$) to respectively $r_c=0.75$ and
$r_c=1.0$, but we preserved a flat $Q=1.2$ at all radii and did not
include a cutout.  The inner slope of the rotation curve in model
Rc1.0 is very slightly below that expected from the disk mass alone,
and so implies a hollow halo at radii $r \la 0.3$, but as the
axisymmetric self-gravity of the disk is ignored, the particles simply
move in the global potential and the disk instability is unaffected by
this undesirable property.

Notice from the top right hand panel of Fig.~\ref{fig.core} that the
logarithmic spiral transform of the mode differs qualitatively from
those from models with smaller $r_c$, although the mode still has a
substantial leading component.  The mode shape in the lower right
panel is more open, having just a single extra antinode near the
center, confirming our suggestion that the steeper inner rise to the
rotation curve is indeed the cause of the tightly-wrapped inner
spiral.  Again, the coherent disturbance continues into the very
centers of these two simulations, indicating that these modes also
reflect off the disk center.

\section{Other results}
\label{sec.others}
Table~\ref{tab.runs} lists the parameter values for most of the
simulations we report here.  It also gives the estimated
eigenfrequency, $\omega$, of the dominant linear mode.  In all
simulations other than the two just presented, we kept the core radius
of the potential, $r_c=0.5$, as in our baseline model.

\begin{figure*}
\includegraphics[width=.56\hsize,angle=270]{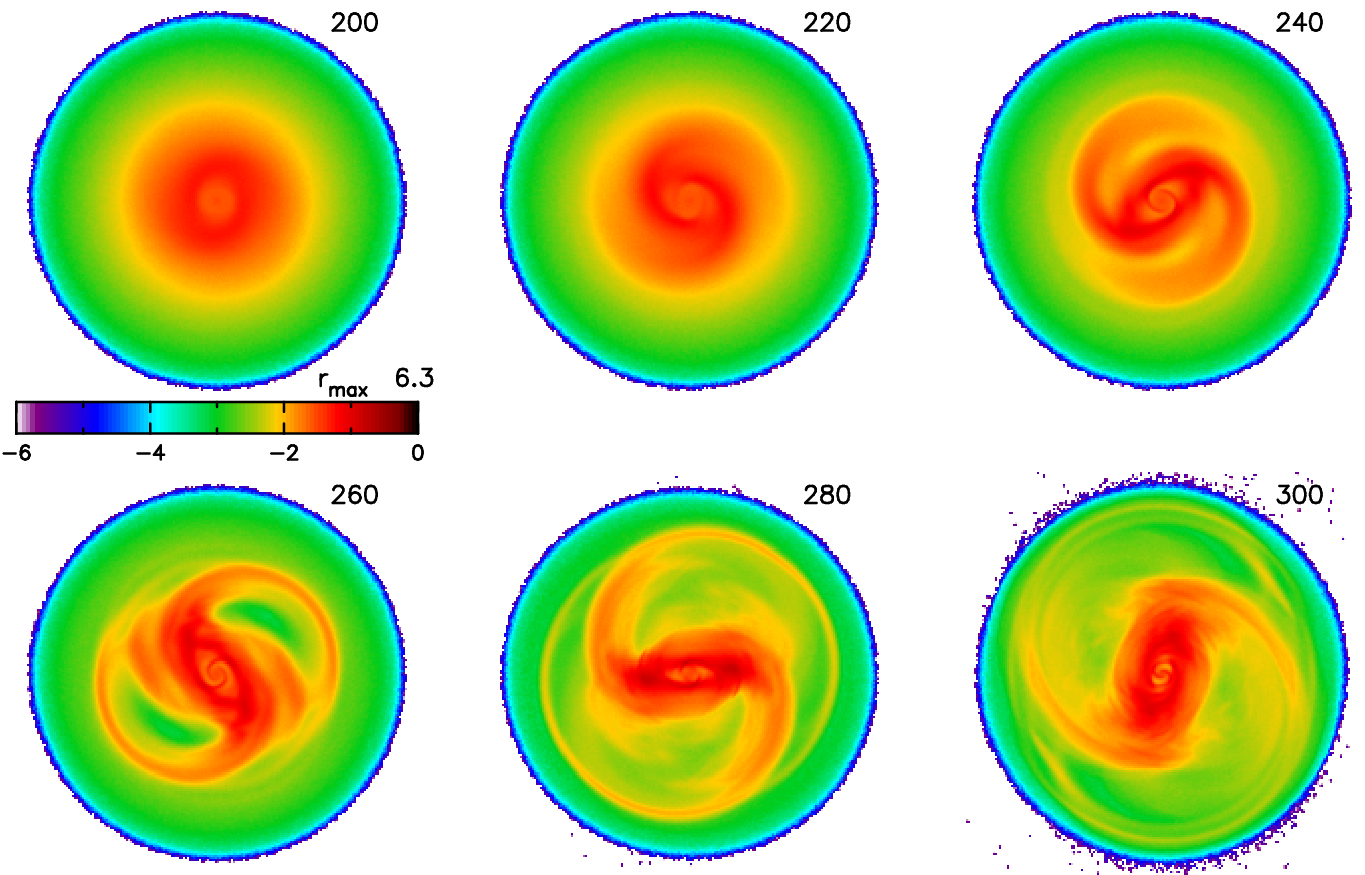}
\caption{The evolution of the disk component in a rerun of 2DC0.2 in
  which self-consistent forces from $m=0$, 2 and 4 all contributed.  A
  large bar formed over the interval $200 \leq t \leq 300$.  Notice
  the inner density minimum, in which a tightly wrapped spiral
  developed, that persists to the last moment shown.  However, the
  model requires a dense bulge (not shown) to support the inner
  rotation curve, which would make inner minimum and disk features
  hard to observe without sophisticated imaging and kinematic data to
  separate disk from bulge stars.}
\label{fig.bar2}
\end{figure*}

\begin{figure*}
\includegraphics[width=.3\hsize,angle=0]{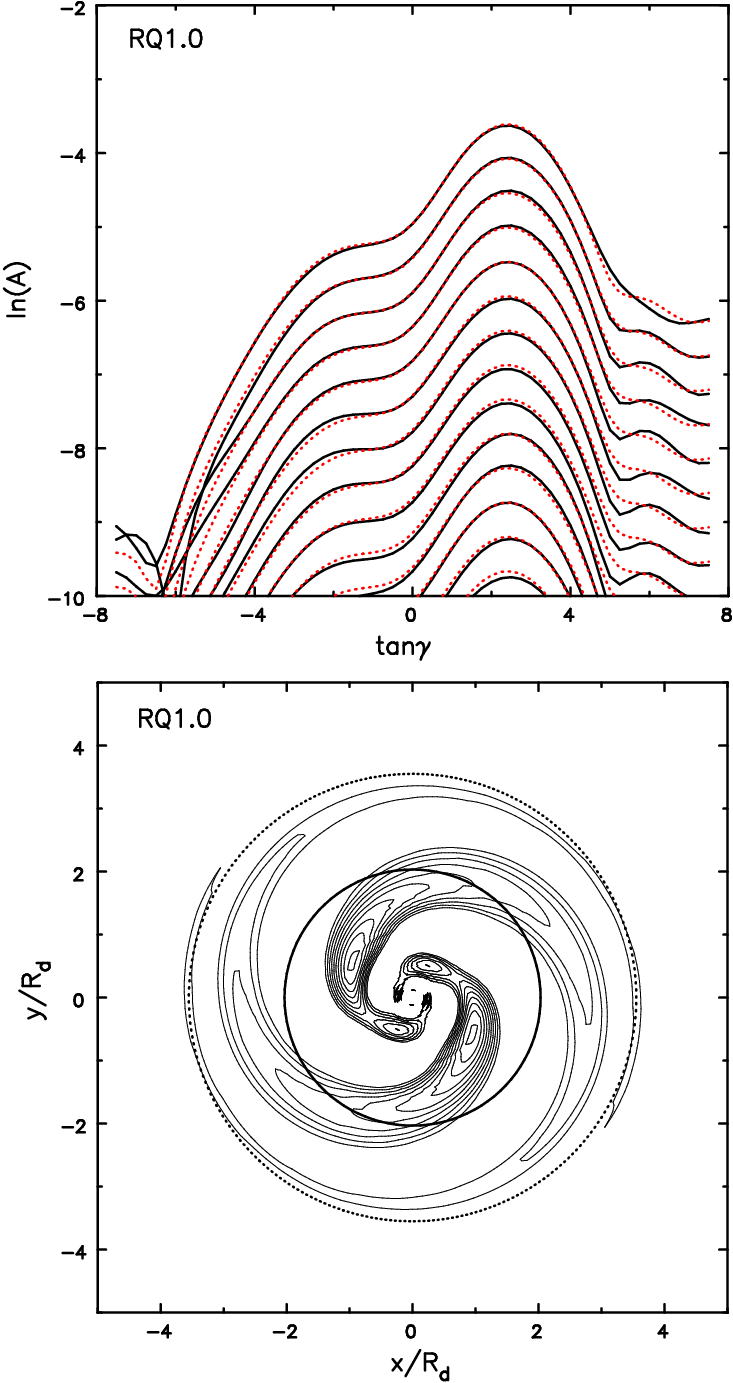}
\includegraphics[width=.3\hsize,angle=0]{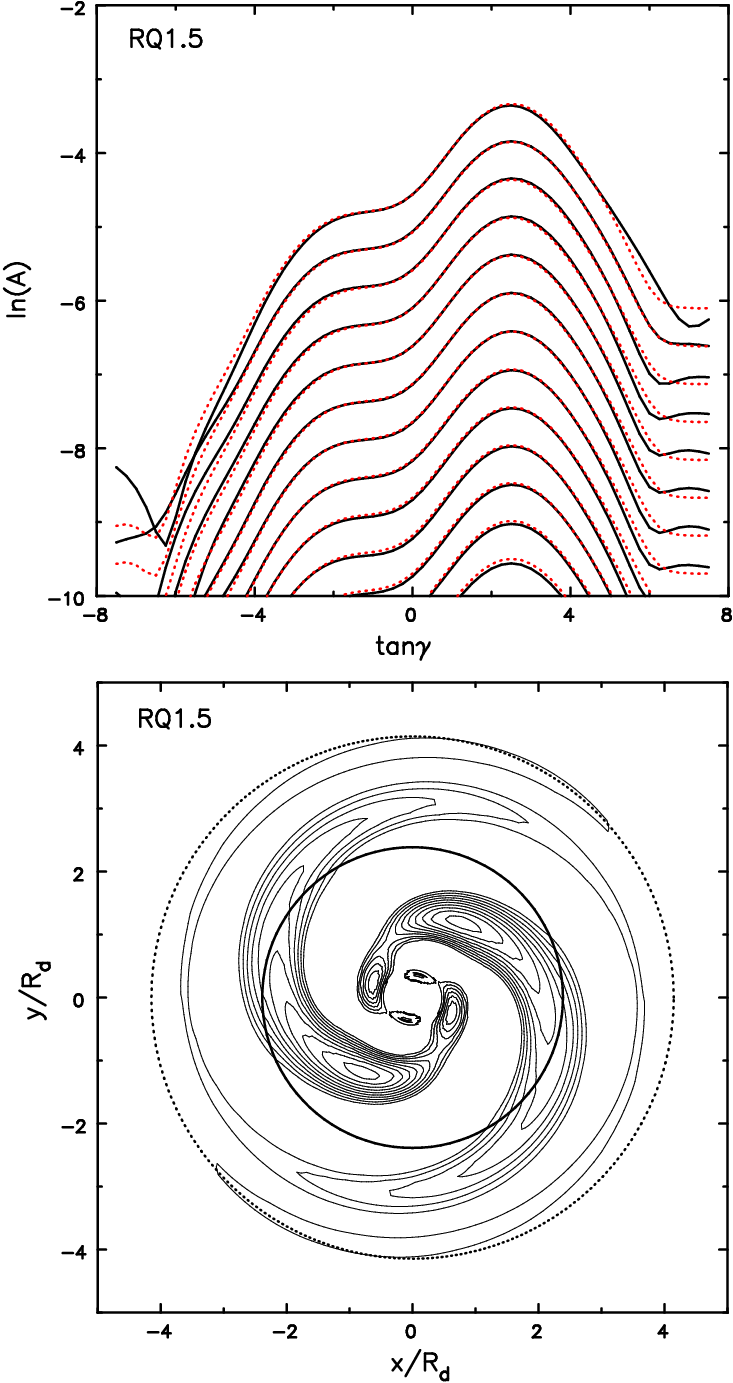}
\includegraphics[width=.3\hsize,angle=0]{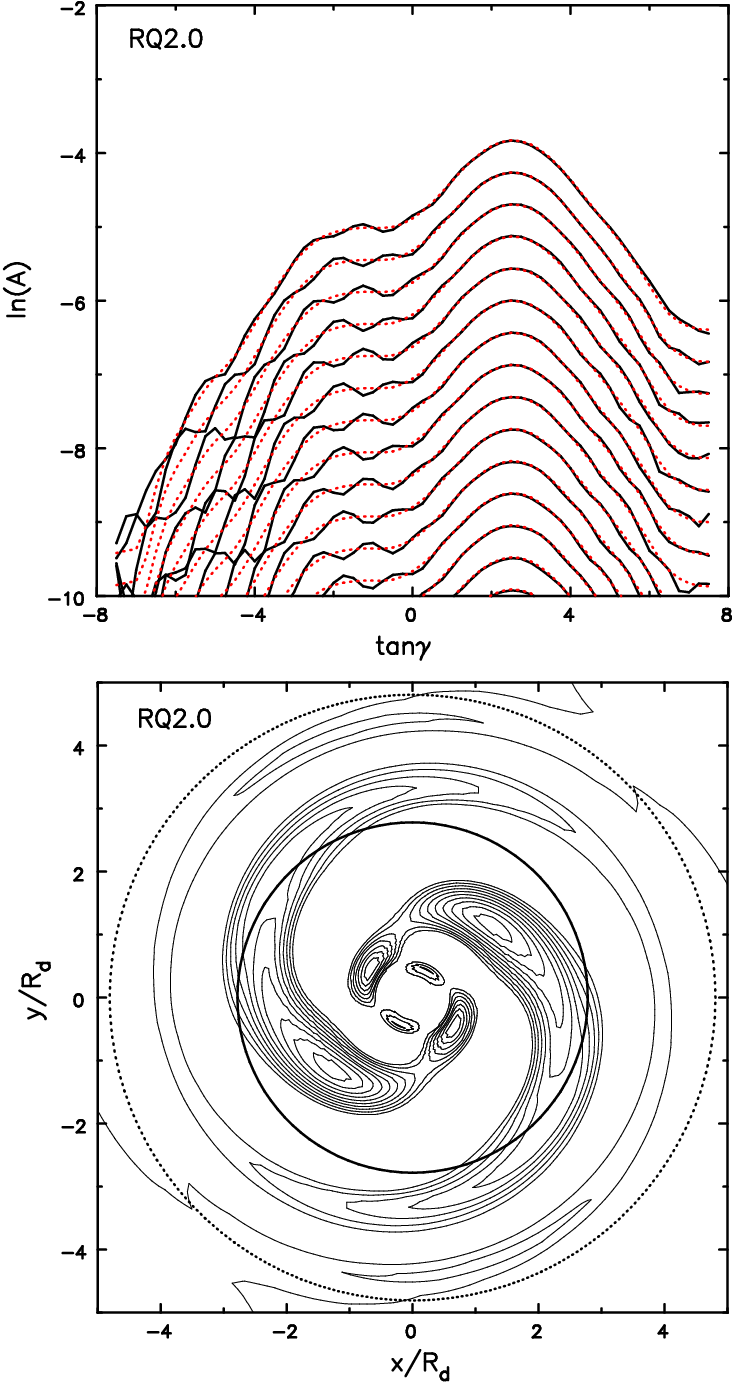}
\caption{Fits to the linear evolution of simulations in which the
  outer radius of the inner density cutout and enhanced $Q$ is varied,
  while $D=0.2$ and $Q(0) = 3$ are held fixed in all three cases.
  Note $R_{\rm cut} = R_Q$.}
\label{fig.varyRcut}
\end{figure*}

The purpose of these simulations is to explore models having higher
$Q$ and/or surface density cutouts in the inner parts of the disk.  It
has proved informative to compare the frequencies, mode transforms,
and mode shapes in sequences of runs from Table~\ref{tab.runs} in
which all but one parameter were held fixed.

\subsection{Disks with hot centers}
We first present a sequence in which we varied the central value of
$Q$, keeping $R_Q=1.5$ and with no disk cutout. We chose $Q(0)=1.0$
(HC1.0) (lower than the baseline model), $Q(0)=1.5$ (HC1.5) and $Q(0)=2.0$
(HC2.0).  The linear modes of these three cases are presented in
Fig.~\ref{fig.risingQ}, and the baseline model fits into this sequence
with $Q(0)=1.2$.

The principal consequence of increasing $Q(0)$ is to slow the pattern
speed, which moves the resonances to large radii and increases the
spatial scale of the mode (lower panels).  Somewhat surprisingly, the
growth rate was little affected as $Q(0)$ increased from 1 to 1.5, but
it decreased when $Q(0)=2$ (Table~\ref{tab.runs}).  Furthermore, the
shapes of the mode transforms in the upper panels all resemble that of
our baseline model, with strong leading components and, despite the
higher $Q$ in the inner disk, the modes all appear to reflect off the
center (lower panels).

\subsection{Cutting away disk mass}
\label{sec.cutout}
Fig.~\ref{fig.varyD} presents results from a sequence of simulations
in which we tapered away increasing quantities of the disk surface
density.  We varied $D=0.8$ to $D=0.2$, keeping $R_{\rm cut}=1.5$ and
maintaining a flat $Q=1.2$ at all radii.  Again, the baseline model
begins this sequence with $D=1.0$.

We have very successfully fitted two modes to the logarithmic spiral
transforms in the upper left panel, but draw only the dominant mode in
the lower left.  The character of the mode transform (upper panels)
differs in the middle and right from that in the left panel, and this
is again reflected most clearly in the mode shape in the bottom right,
which manifestly does not extend into the center -- we have vainly
searched for any coherent waves in the region $R<0.5$ in 1DC0.2.  The
inner edge of the mode in 1DC0.5 (middle panel) is not as decisively
outside the center, but we could find little evidence for a coherent
wave inside $R<0.4$.  Also we do not find a clear trend in the pattern
speeds as the cutout is deepened, but the vertical spacing of the
curves in the upper panel gives a clear visual indication that the
growth rates of the dominant mode rises from left to right, as
confirmed by the numerical values in Table~\ref{tab.runs}.

One consequence of cutting away mass from the inner disk is that the
denominator of eq.~(\ref{eq.wavenumber}) contains the factor $\Sigma$,
suggesting that $k$ should increase and spiral modes would become
still more tightly wrapped.  We do not see this happen in
Fig.~\ref{fig.varyD}, but instead the disk center, where the mode
would be most tightly wrapped, does not appear to support a
disturbance at all.  We address this result in \S\ref{sec.innref}.

Fig.~\ref{fig.varyD2} presents a parallel sequence to
Fig.~\ref{fig.varyD}, but differs because the central value of $Q$ was
also raised to $Q(0)=2$.  We illustrate just three cases from this
sequence for which $D=0.8$ (2DC0.8), 0.6 (2DC0.6), and 0.4 (2DC0.4),
but the sequence includes $D=1$ (HC2.0) (right panel of
Fig.~\ref{fig.risingQ}) and $D=0.2$ (2DC0.2) (illustrated below).

Fig.~\ref{fig.bar2} displays the last part of the evolution of a rerun
of simulation 2DC0.2 to show how a deep inner cutout affects the
non-linear formation of a bar.  In addition to the usual $m=2$
disturbance forces, this simulation included the self-consistent
axisymmetric term ($m=0$) of the disk particles as well as the $m=4$
term, in order to capture their influence on the non-linear evolution.
Note, the inclusion of the $m=4$ term required us to increase the
number of particles on each half-ring from 3 to 5 in order to maintain
a quiet start.  The extra force terms have no effect on the $m=2$ mode
frequency or shape, as expected from theory, and we were unable to
detect an instability at $m=4$.  The limiting amplitude of the bar is
increased over that in the $m=2$ only case and the resulting bar,
which has a density minimum at its center, is a little thinner.  In
reality, the density minimum would probably be obsured by a dense
bulge.

\subsection{Varying the radial extent of the cutout}
Fig.~\ref{fig.varyRcut} shows the consequence of varying the radial
extent of the cutout.  In all three cases, $Q(0)=3$ and $D=0.2$,
but $R_{\rm cut}=1.0$ in RQ1.0 (left), $R_{\rm cut}=1.5$ in RQ1.5 (middle)
and $R_{\rm cut}=2.0$ in RQ2.0 (right).

We found from the modes in Fig~\ref{fig.varyD} that a deep central
cutout out erases the multiple peaks in the mode transform for the
baseline model (top panel of Fig~\ref{fig.example}) and similar cases,
as it appears to have done here, but here we have also raised the
central $Q(0)$.  Despite the simple appearance of the mode transform,
the modes in Fig.~\ref{fig.varyRcut} are still quite tightly wrapped,
though only in the cool ($Q=1.2$) outer disk beyond $R_Q = R_{\rm
  cut}$.  The growth rate of the dominant mode is little affected by
increasing $R_{\rm cut}$, but the pattern speed is reduced and, while
the spatial scale of the mode rises as a consequence, it appears to do
so roughly homologously.  Note that none of these modes extends into
the center, and we find that the inner edge of the mode moves to
larger radius as the radial extent of the cutout is increased.

\begin{figure}
\begin{center}
\includegraphics[width=.8\hsize,angle=0]{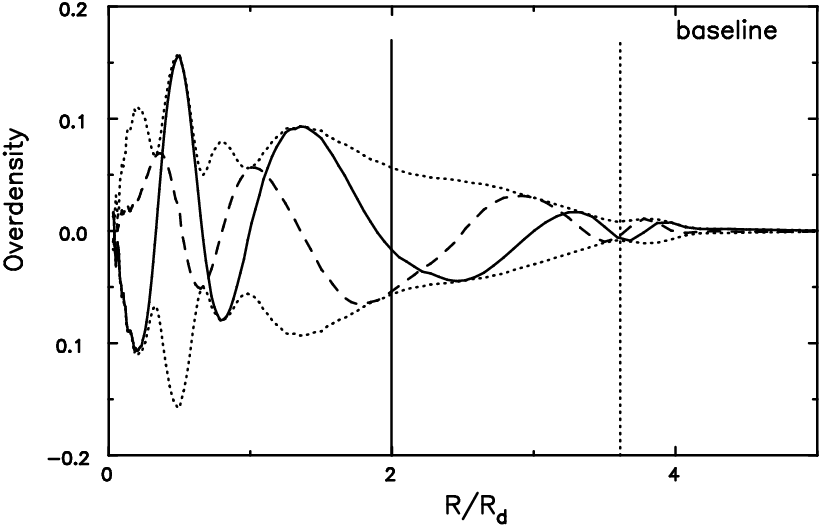}
\end{center}
\caption{The radial variation of the fitted mode amplitude in our
  baseline model.  The solid curve shows $\Re\{A_m(R)\}$, the dashed
  curve $\Im\{A_m(R)\}$, while the dotted curves mark the mode
  envelope $\pm|A_m(R)|$.  The vertical lines mark the radii of the
  resonances: CR solid line and OLR dotted.}
\label{fig.rdbasic}
\end{figure}

\section{Modes in cutout disks}
\label{sec.innref}
We reported in \S\ref{sec.example} that the mode in our baseline model
reflected off the center. Fig.~\ref{fig.rdbasic} presents additional
evidence in support this statement because the mode amplitude indeed
remains significant to small radii and declines steeply to $R=0$.

\begin{figure}
\begin{center}
\includegraphics[width=.8\hsize,angle=0]{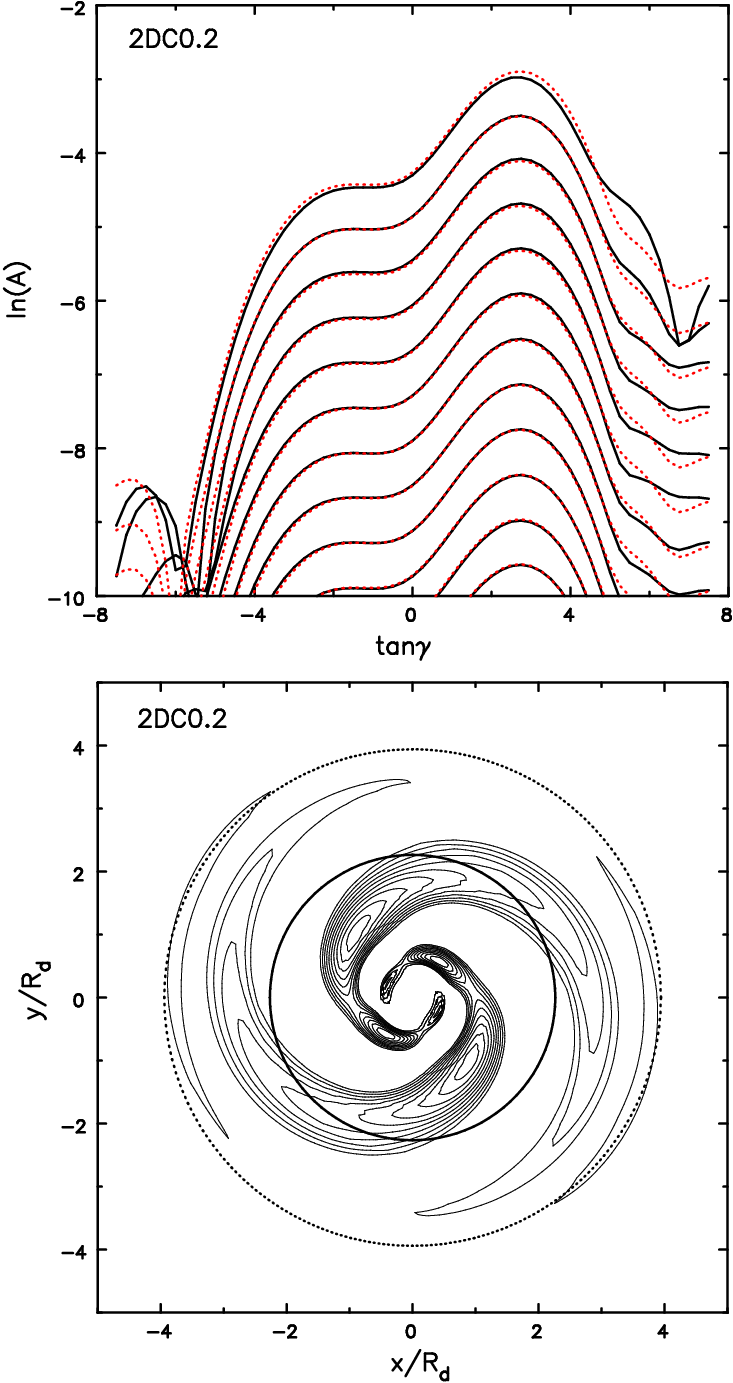}
\includegraphics[width=.8\hsize,angle=0]{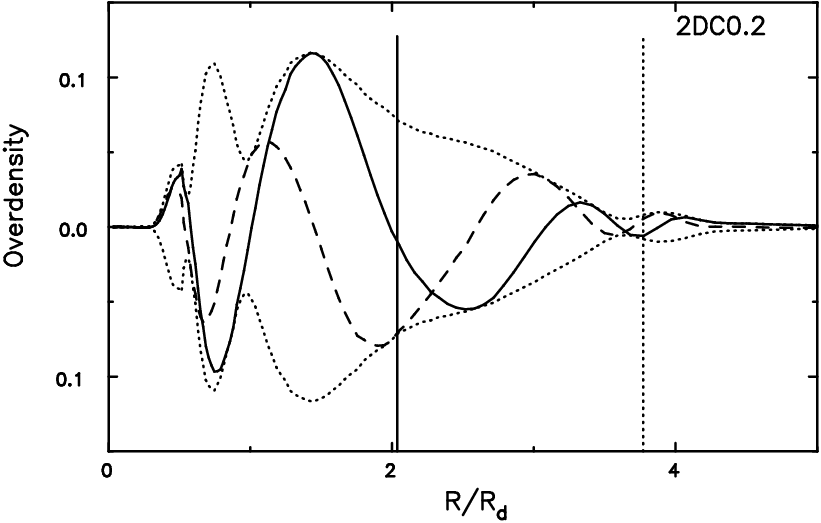}
\end{center}
\caption{The linear mode of run 2DC0.2, which is the continuation of
  the sequence shown in Fig.~\ref{fig.varyD2} to the case for which
  $D=0.2$.  As in the rest of the sequence, $Q(0)=2$ and $R_{\rm cut}
  = R_Q = 1.5$.  The bottom panel shows the radial variation of the
  fitted mode amplitude, as described for Fig~\ref{fig.rdbasic}.  Note
  that the mode amplitude is flat and near zero for $R \la 0.3$. }
\label{fig.run5315}
\end{figure}

However, in some of the cases reported in \S\ref{sec.others}, it
seemed that the mode amplitude dropped to near zero inside a finite
radius, and we here show another particularly clear example in
Fig.~\ref{fig.run5315}.  This model, 2DC0.2, continues the sequence
shown in Fig.~\ref{fig.varyD2} to the case for which $D=0.2$, and has
no coherent wave at radii $R \la 0.3$.  We have added the bottom panel
confirming that the amplitude inside this radius is $ \ll 1\%$ of the
peak and seems consistent with noise, which implies that the
disturbance is confined to radii $R \ga 0.3$.  Note that the mode
still possesses a strong leading component (top panel of
Fig.~\ref{fig.run5315}).

\subsection{Numerical checks}
We have checked carefully that the inner edge of the mode it is not a
numerical artifact.  It seemed possible that representation of a very
tightly wrapped trailing wave on a polar grid may alias as a similarly
tightly wrapped leading wave, causing the disturbance to reflect off
the grid.  In order to check that this is not the case, we reran the
same initial file of particles in two additional simulations in which
we first doubled and then halved the number of grid points in each
dimension.  These changes will have, respectively, halved and doubled
the spacing of the grid rings, but neither the inner edge of the mode,
as shown in the left panel of Fig.~\ref{fig.checks}, nor the mode
frequency changed thereby ruling out reflection due to grid aliasing.

\setbox1=\vbox{\hbox{
\includegraphics[width=.45\hsize,angle=0]{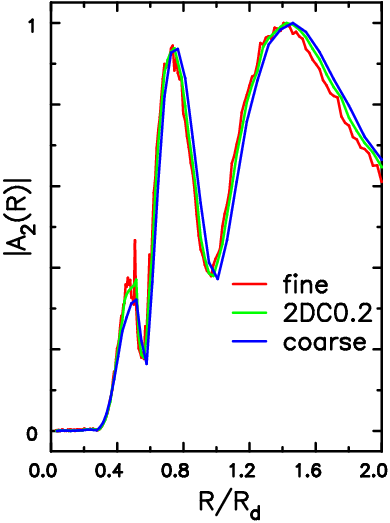} \quad
\includegraphics[width=.45\hsize,angle=0]{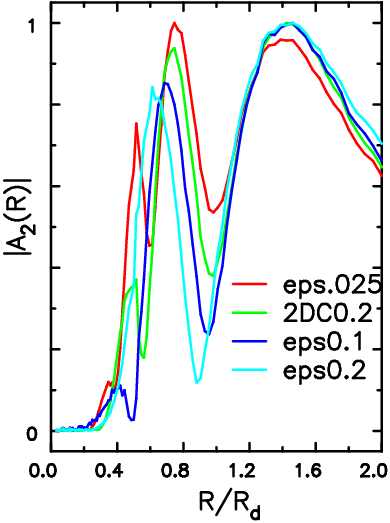}
}}
\begin{figure}
\begin{center}
  \box1
\end{center}
\caption{The radial variation of the inner part of $|A_2(R)|$ of the
  fitted mode to tests of the numerical parameters in simulations with
  the same initial file of particles used for run 2DC0.2
  (Fig~\ref{fig.run5315}).  These simulation tests are not listed in
  Table~\ref{tab.runs}.  All curves were rescaled to span $0 \leq
  |A_2(R)| \leq 1$. Left panel: the red and blue lines are from cases
  where the resolution was respectively doubled and halved from our
  standard grid, which is reproduced as the green line.  Right panel:
  the red, blue and cyan lines are from cases in which $\epsilon$ was
  respectively halved, and increased by factors of two and four from
  our standard value, which is reproduced as the green line.}
\label{fig.checks}
\end{figure}

Another possibility is that gravity softening may be playing a role,
since it increasingly attenuates the self-gravity of density waves by
the factor $e^{-|k|\epsilon}$, and may prevent them from propagating
at all when the wavelength ($=2\pi/|k|$) decreases below some
possible minimum value.  We have therefore tried both halving and
increasing the softening length by factors of two and four.  Note that
a change to the softening length affects the entire mode, changing
both parts of the eigenfrequency, the positions of the resonances, and
the radial wavelength of the mode.  However, the mode in each case
again had tiny amplitude at radii $R<0.2$, though in two cases it was
small but greater than zero between $0.2<R<0.3$.  We have also
verified that halving the time step made no difference to the mode
shape or frequency.  These successful checks indicate that the mode
not extending into the center is physically real.

\begin{figure}
\begin{center}
\includegraphics[width=.8\hsize,angle=0]{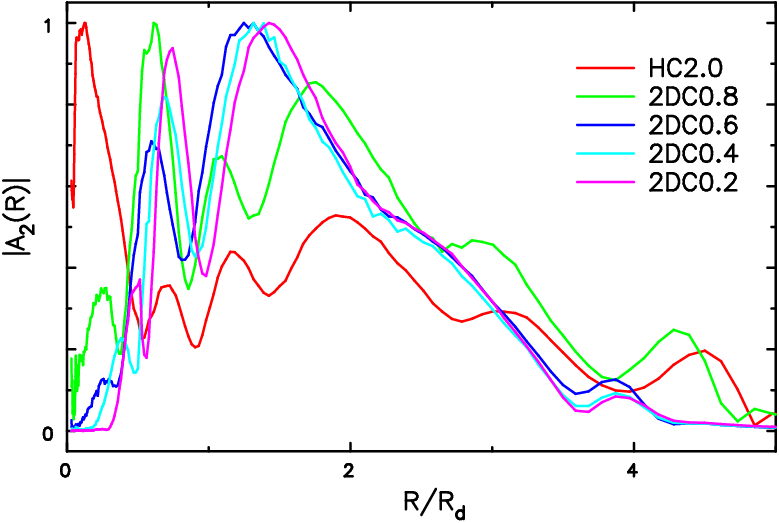}
\includegraphics[width=.8\hsize,angle=0]{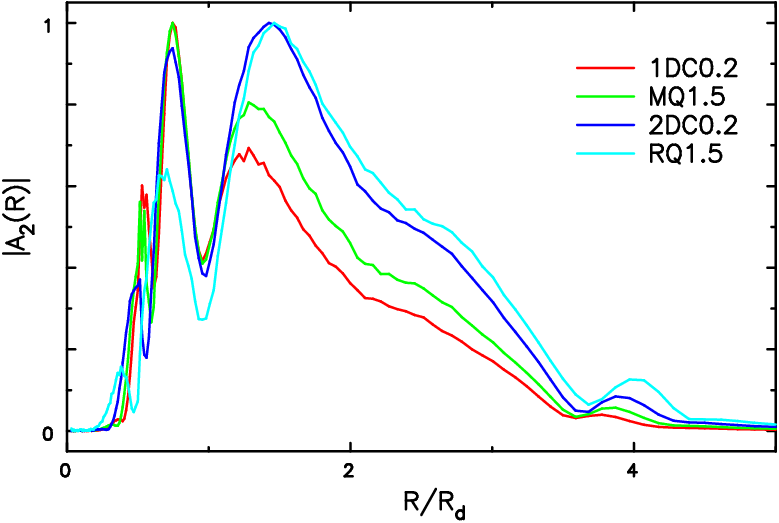}
\end{center}
\caption{The radial variation of $|A_2(R)|$ of the fitted mode as the
  physical parameters were changed.  Again all curves were rescaled to
  span $0 \leq |A_2(R)| \leq 1$.  The lines in the upper panel are
  from the indicated runs in which $1 \geq D \geq 0.2$ in steps of 0.2
  while $R_{\rm cut}=1.5$ and $Q(0)=2$ are fixed. Notice that the mode
  extends into the center when $D \geq 0.6$, but appears not to when
  $D=0.4$ (cyan) and $D=0.2$ (magenta).  The lines in the lower panel
  are from runs in which $R_{\rm cut}=1.5$ and $D=0.2$ while $Q(0)$ is
  increased from $Q(0) = 1.2$ to $Q(0)= 3$.  It is remarkable that the
  central value of $Q$ has such a small effect on the inner edge of
  the mode.}
\label{fig.varyQ0}
\end{figure}

\subsection{Summary of the evidence}
We report in Fig.~\ref{fig.varyQ0} the radial amplitudes of fitted
eigenmodes as the physical parameters $D$ and $Q(0)$ are varied.  The
sequence in the top panel indicates that the modes do not reach the
center when $D \geq 0.5$, a finding that seems to hold in all
simulations reported in Table~\ref{tab.runs}, irrespective of either
$R_{\rm cut}$ or $Q(0)$.

In fact, the evidence in the lower panel of Fig.~\ref{fig.varyQ0},
which shows a sequence with increasing $Q(0)$, indicates that the
inner radius of the mode is little affected by increasing random
motion.  It is true that the inner radius of mode does move inwards
(slightly) as $Q$ rises -- the cyan line is from the model with the
hottest center, the red from the coolest.  However, the important
density gradient is probably that of the guiding centers, which is
approximately unchanged in this sequence of models, and the sharp
inner edge of the mode in the cooler disks is increasingly blurred by
larger epicycles in the hotter disks.  Note also that we reported in
the right panel of Fig.~\ref{fig.checks} that the inner radius of the
mode is also little affected by the value of $\epsilon$.

\subsection{Mode mechanism}
Since it was provoked by a steep inner density gradient, it seemed
natural to suppose that the inner mode was related to the better-known
outer edge modes of disks \citep{Toom81, PL89}.  Outer edge modes
occur where mild non-axisymmetric distortions in a steep surface
density gradient create co-orbiting over-densities.  Each over-density
induces a swing amplified \citep{JT66, Binn20} supporting wake in the
interior disk that gives angular momentum to the edge overdensity,
causing it to rise outwards, and therefore the edge distortion grows.

Thus, a possible inner edge mode could operate by a similar mechanism.
Co-orbiting overdensities could be created by mild non-axisymmetric
distortions at the radius of steepest density gradient that would
induce a strong trailing response in the higher surface density
region.  The attraction from this wake would cause the overdensities
to be pulled back and therefore to sink towards the rotation center,
creating a similar instability.  However, it cannot be the mechanism
of the modes we report from our simulations, because then the
corotation radius of the modes would be close to the radius of the
steepest gradient, which it manifestly is not in
Figs.~\ref{fig.varyD}, \ref{fig.varyD2}, \ref{fig.varyRcut}, and
\ref{fig.run5315}.

Similar inner modes were reported by \citet[][see also \citealt{ER98,
    SE01}]{Zang76}, who found instabilities in the Mestel disk when
the center was cut out too sharply.  In his case also, the slow
pattern speed of the $m=2$ mode placed corotation at more than twice
the radius of the steepest surface density gradient, as is the case
for the modes we find.  Zang noted in his thesis ``Although we have
some ideas about the details of this process, we are presently unable
to offer a \underline{clear} ({\it sic}) physical mechanism'' for the
instability provoked by the sharp inner cutout.  \citet{ER98} remark
``If, however, the cut-out is sufficiently sharp (as for larger
cut-out indices), it presents a barrier which reflects the incoming
trailing waves.''  Thus they clearly view the mode as a slightly
different type of cavity mode, but they do not offer any further
explanation for the nature of the barrier or the reflection.

The strength of the leading wave component in every one of the modes
in our simulations is clear evidence that swing amplification plays a
major role, even when the mode does not reach the center.  As
explained above, an outer edge mode does not require a feedback loop,
because the instability is excited at corotation, but it is hard to
see how the slow modes, found by \citet{Zang76}, \citet{ER98}, and
here in this paper, could grow without feedback.  An inner reflection
off the center is excluded for $m\geq2$ waves in cusped potentials,
and we observe that the modes we find in cutout disks do not reach the
center either.  So we concur with \citet{ER98} that an inner
reflection must occur at a finite radius in order to turn inwardly
propagating trailing waves into outwardly propagating leading waves.

\begin{figure}
\begin{center}
\includegraphics[width=.9\hsize,angle=0]{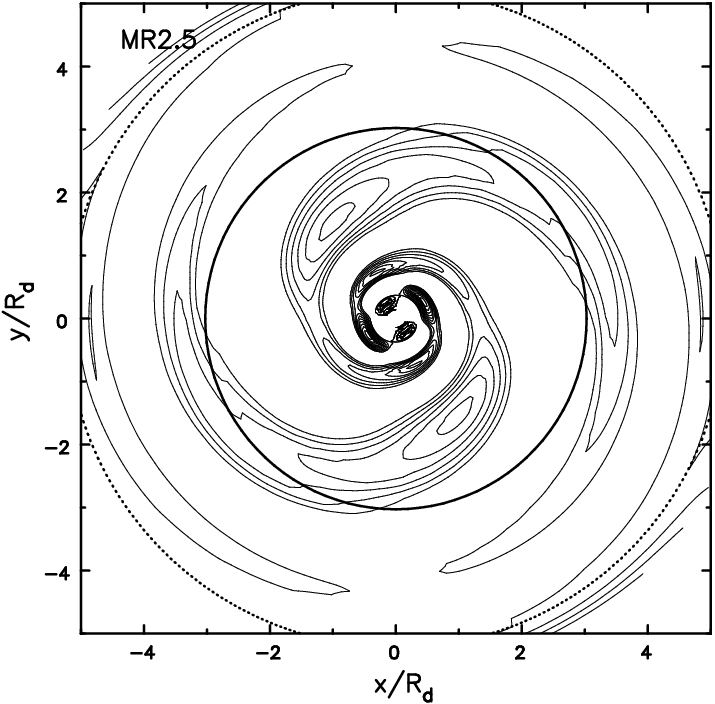}
\includegraphics[width=.9\hsize,angle=0]{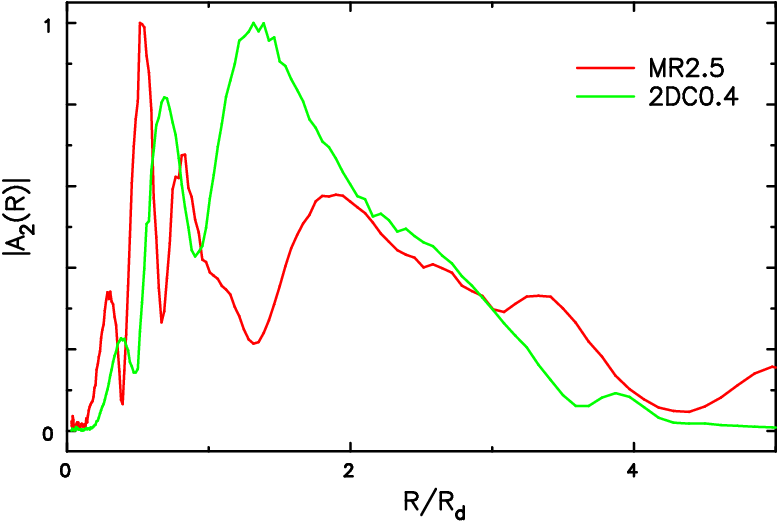}
\end{center}
\caption{Above: the fitted mode of model MR2.5.  Below: comparison
  between the mode amplitude profiles as Rcut is increased.  Note that
  the inner reflection is closer to the center in MR2.5 than in
  2DC0.4, for which $R_{\rm cut}=1.5$.}
\label{fig.MR2.5}
\end{figure}

\subsection{Possible reflection mechanism}
We offer the following speculative idea for how an inner reflection
could happen.  WKB wave theory (eq.~\ref{eq.wavenumber}) suggests that
spiral waves should become more tightly wrapped in models having
deeper surface density cutouts.  Thus inwardly travelling waves will
wrap ever more tightly as they enter the cutout region.  As this
happens, the ability of a disk with random motion to carry the wave
must dwindle because stars whose epicycle radii are greater than the
diminishing wavelength of the wave cannot provide much of a supporting
response.  A wave supported by an ever decreasing fraction of stars in
the center of the velocity distribution cannot sustain its amplitude
as $|k|$ rises.  The collisionless nature of a stellar disk excludes
possible dissipation and, if there is no nearby resonance where
wave-particle interactions can occur \citep{LBK}, wave action must be
conserved and the only possible outcome is that the wave must bounce.

To follow up this idea, we have attempted to use the measured pattern
speeds from Table~\ref{tab.runs} to solve eq.~(\ref{eq.WKB}) for
$k(R)$, including the full functional form for $\cal F$ as well as a
factor $e^{-|k|\epsilon}$ that further attenuates the gravity term.
Unfortunately, setting $Q(R)$ from eq.~(\ref{eq.Qprof}), together with
the gravity softening term, prevented us from finding solutions for
$|k|$ over almost the entire radial range inside the CR, because the
forbidden region, an artifact of eq.~(\ref{eq.WKB}) that stems from
assuming a steady wave, is so broad.  We were therefore unable to make
even a very rough estimate of radius at which the wave would bounce,
which anyway would be inconsistent with the infinite plane wave
approximation that underlies eq.~(\ref{eq.WKB}).

Not only is the idea that the wave might reflect off the decreasing
density gradient highly speculative, but it is also hard to reconcile
with the evidence in the bottom panel of Fig.~\ref{fig.varyQ0}, which
indicates that the reflection radius is little changed as the central
$Q$ is increased from 1.2 to 3, with everything else held fixed; the
slight change could be consistent with epicyclic blurring, as we note
above.  It seems unlikely that the reflection radius would be
independent of the degree of random motion, since $|k|$ for steady
waves is strongly affected by $Q$, but we have no other idea to offer.

\subsection{Gentler inner tapers}
Since the instability of disks in cusped potentials disappears when
the taper is made more gentle, it is natural to ask whether the same
happens in our case.  Our inner taper function (eq.~\ref{eq.Sigtap})
differs from that used by \citet{Zang76} and \citet{ER98}, but we can
make it more gentle by increasing $R_{\rm cut}$.  We therefore present
two more simulations, listed as the last two entries in
Table~\ref{tab.runs}.  These two models differ in just one respect
from 2DC0.4, reported in the right hand panels of
Fig.~\ref{fig.varyD2}: $R_{\rm cut} = 2.5$ in model MR2.5 and $R_{\rm
  cut} = 3.5$ in model MR3.5.

The upper panel of Fig.\ref{fig.MR2.5} illustrates the slower and
milder instability, in comparison with that in 2DC0.4, we obtained
from simulation MR2.5.  Notice from the lower panel that the
reflection radius is a little closer to the center than that in model
2DC0.4 even though $R_{\rm cut}$ has been increased.  We could not
detect a growing disturbance in MR3.5.  Thus we do find that gentler
inner tapers are stabilizing, as did \citet{Zang76} and \citet{ER98},
but we believe it is for a different reason, as follows.

\begin{figure}
\begin{center}
\includegraphics[width=.9\hsize,angle=0]{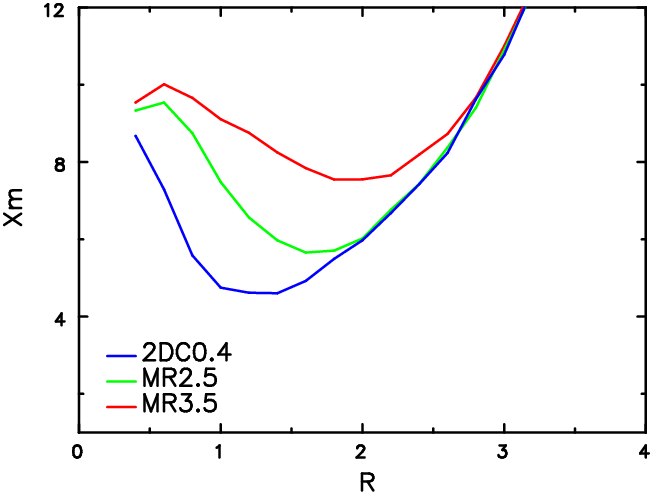}
\end{center}
\caption{The function $Xm = 2\pi R/\lambda_{\rm crit}$ for the indicated
  three models.  Since $m=2$, we see that the swing-amplification
  parameter $X>3$ at all radii for the stable model MR3.5.}
\label{fig.Xm}
\end{figure}

The $m=2$ instability of our baseline model is a standing wave between
the center and corotation where it amplifies, as in the usual
bar-mode picture.  When we apply a deep cut out, we find that
reflection can occur away from the center, but we still have a cavity
mode with amplification at corotation.

\citet{Zang76} and \citet{ER98}, on the other hand, calculated disk
modes in a cusped potential, which precludes feedback through the
center (except for $m=1$), finding that models having gentle cutouts
had no global instabilities for $m\geq2$.  They also found that sharp
inner cutouts provoked instabilities, and \citet{ER98} asserted that
swing amplification farther out in the disk could combine with
reflection off a sharp cutout to allow a cavity mode.  So the modes in
sharply cutout models, in both their case and in ours, apparently have
the same mechanism.

However, when we make a gentler inner cutout by increasing $R_{\rm
  cut}$ at a fixed central depth, we also cut away a lot more mass
from the disk, which eventually turns off $m=2$ swing amplification
because $X>3$ over the whole disk, as shown in Fig.~\ref{fig.Xm}.  The
different taper function in their otherwise smooth, self-similar disks
leaves the swing amplifier still eager to operate in the outer disk,
but it is denied feedback because the ILR absorbs infinitesimal
incoming waves when the inner taper does not make a barrier.  Thus the
explanation for stability in cusped models is quite different: the
stability of MR3.5 results from no other physical mechanism than that
argued by \citet{ELN}, who found that increased halo crushes $m=2$
swing amplified waves!

\section{Discussion}
\label{sec.discuss}
The simulations presented in this paper capture the important
gravitational dynamics of the disk, but are otherwise highly
idealized, with several simplifying assumptions summarized in
\S\ref{sec.assumps}, whereas the real universe is far more
complicated.  In the light of this consideration, and at the
suggestion of the referee, we here re-examine our assumptions and
findings from isolated simulations.

\subsection{The case of M33}
The local group galaxy, M33, is perhaps the poster child of the bar
stability problem.

\citet{SSL} closely examined the stability of M33 in a set of models
that matched all available data, but were unable to account for the
absence of a bar in that well-studied galaxy.  Their fully
self-consistent 3D simulations with live halos modeled the gas in a
variety of ways, none of which prevented the rapid formation of a bar.
The stellar and gaseous density profiles of the disk were constrained
by observations, which exclude possible cut outs and, although the
stellar velocity dispersion is known only in the center, that is the
most important constraint.  The regular gas kinematics of the inner
disk \citep{Corb14, Kam17} indicates that it has not been disturbed by
infalling substructure in the recent past and the pronounced warp is
sufficiently far out to have no plausible effect on the stability of
the inner disk.  While all their models had smooth halos, some rotated
both with, or counter to, the rotation sense of the disk.  Reducing
the stellar mass-to-light ratio to some quite unreasonably low value
did lead to milder instabilities, but the disk then supported
multi-arm spirals, whereas near IR images of the galaxy
\citep[\eg][]{Kam15} indicate two major spiral arms with perhaps a
weaker third arm.  Their attempts to induce a open bi-symmetric spiral
in a lower mass disk through a tidal interaction were not successful.

\citet{Smer23} posted a study of star counts in M33 which indicated a
bar-like structure in the inner 1kpc of the disk.  We note that the
strong bar created by global instability in the simulations by
\citet{SSL} had a semi-major axis of $\sim\;3$kpc, and therefore the
finding of a short, weak bar in M33 does not alter the conclusion that
the apparent global stability of M33 presents a dynamical challenge.

Thus the study by \citet{SSL} did not make many of the simplifying
assumptions listed in \S\ref{sec.assumps}, yet all their plausible
models still formed a strong bar within 1\;Gyr.  A possible weakness
of their study is that the halos lacked the substructure expected in a
hierarchical universe.  However, massive subhalos would disrupt the
disk, which clearly has not happened recently in M33, while low-mass
subhalos would have little effect on disk stability and anyway
probably would not survive at all in the inner halo
\citep[\eg][]{Sawa17}.

\subsection{The present study}
Our purpose in this present study was to revert to highly idealized
simulations in which the dynamics is firmly under the control of the
experimenter, to determine whether the properties of the disk center
could be changed in such a way as to inhibit the bar instability.
Although we made a number of simplifications, listed in
\S\ref{sec.assumps}, in order to obtain reliable results at low
computional cost, those approximations have been examined before and
most are known to have little or no effect on the stability of the
entire disk; the exception being the assumption of a rigid halo that,
when relaxed, is known to enhance the growth rate of the bar mode, as
noted in the introduction.

Unfortunately, we have found that neither deep cut outs nor a high
central $Q$ have much effect on the global stability of the models we
have tried.  These models can be thought of as more closely
corresponding to galaxies, such as the Milky Way, that have higher
mass than does M33, though we made no attempt to model any particular
galaxy, neither do they bear a close resemblance to those created in
cosmological simulations.  It is therefore reasonable to ask whether
more realistic models would yield a different result?  Our baseline
model adopts an exponential disk and a cored isothermal potential.

It seems unlikely to us that a different disk mass profile would make
a qualitative difference to global stability.  We base this
expectation on Toomre's mechanism for the bar mode, which requires
only swing amplification at corotation and an inner reflection of
inwardly traveling trailing waves into outwardly traveling leading
waves.  There seems little reason to expect that the modes of any
other disk mass profile would differ fundamentally, as long the waves
are able to propagate inside corotation and the disk is heavy enough
to amplify $m=2$ disturbances.  Indeed, as reviewed in the
introduction, we have evidence in the literature that the isochrone,
Kuzmin-Toomre, and Gaussian disks all possess bar modes of the type we
are exploring.  Note also that our paper already explored disks of
other mass profiles, because we changed the baseline exponential
profile by cut outs of various depths and extents.

The cored isothermal potential is perhaps more questionable, since we
know from the studies of \citet{Zang76} and of \citet{ER98} that disks
in cusped potentials can be {\em linearly} stable to all modes of
$m\geq2$, as long as there are no sharp edges or grooves in the disk.
The global stability of such disks is precarious, however, because
even quite mild non-linear effects can lead eventually to a strong bar
\citep[\eg][]{Sell12}.  We chose a cored model so as not to repeat
these earlier studies.  Cosmological simulations could also motivate
the choice of a cusped halo mass profile, but numerous studies of
galaxies have questioned whether real halos are cusped, with the
balance of the evidence favoring cores \citep[\eg][]{Wein15, LiPe20}.
Our cored isothermal potential also implies a halo density profile
that drops as $r^{-2}$ at large radii, which is slower than the
$r^{-3}$ power expected from cosmologically simulated halos, but this
cannot affect disk stability because spherically distributed matter
well outside the disk exerts no forces on the interior.  Thus we doubt
that our conclusion that deep disk cut outs and/or hot disk centers
have little effect on global stability is dependent on our adopted
potential.

Of course, this simple analytic potential is perfectly spherical,
smooth, and unresponsive -- properties that are not expected for real
halos formed in a hierarchical universe.  However, we might expect
that a lumpy halo would promote disk instability, since
non-axisymmetric disturbances in the disk induced by passing halo
inhomogeneities would be swing-amplified and thereby contribute to the
growth of bars.  Also, as noted above, a halo composed of mobile
particles is able to enhance bar growth, so strong instabilities in
our rigid halo would grow yet more rapidly in a live one, which does
not help the problem at hand.  Note also that halo rotation does
affect bar growth \citep{SN13, Sell15, Coll19}, but has not so far
been found to stabilize the disk.  These arguments do not, however,
exclude an, as yet unknown, factor that may allow unbarred disks to
survive in cosmological simulations.

\section{Conclusions}
\label{sec.concs}
All models but the last (MR3.5) presented here have turned out to be
moderately to strongly unstable, and we have found little evidence
from this study that surface density cut outs or hot centers
contribute to disk stability.  This finding is independent of changes
to grid resolution, number of particles, time step, or grid
geometry.\footnote{Mode growth rates are reduced by increasing gravity
softening, because forces from density disturbances are weakened, but
even unreasonably large values of $\epsilon$ are unable entirely to
suppress a bar instability \citep{Erick74}.}

However, the unstable mode is not always due to a standing wave in a
cavity between the center and corotation, with amplification at
corotation in the manner described by \citet{Toom81}.  The modes in
deeply cutout disks reflect outside the center, although these
slightly different instabilities also saturate as strong bars.  In
both cases, the mode transforms include substantial leading components
that evolve to amplified trailing waves, causing the net overall
trailing spiral appearance for the mode.  We therefore conclude that
both types of mode rely on swing amplification, which is vigorous for
bisymmetric waves in the somewhat heavy disks we adopt.

Our adopted rotation curve rises sufficiently gently that ILRs can be
avoided enabling the cavity mode cycle described by \citet{Toom81} in
disks that do not have deep density cutouts.  But those modes generally
had a more tightly-wrapped spiral shape in the inner disk than is
usual for bar instabilities, and we demonstrated this was because the
unresponsive mass in a bulge or inner halo, required by the adopted
rotation curve, raises $\kappa$ and decreases the preferred wavelength
in the inner disk.

Our objective in this study was to test whether the bar instability
could be quelled by changing the properties of the inner disk to
impede the transmission of waves through the center.  However, we have
found that deep inner cutouts in the disk are not stabilizing because
the cavity mode in this case reflects off the sharp cutout, and we
have shown that this once again leads to a strong bar.  We tried more
gentle cutouts, which did eventually inhibit the bar-instability, but
only because the disk mass was so drastically reduced by the extensive
cutout that swing-amplification of $m=2$ vaves was inhibited at all
radii.  It is, however, rather surprising that disks having hot
centers are also quite unstable.

Although our models do not exactly match those studied by
\citet{BLLT}, we do not reproduce their predicted more mild
instabilities in a disk having a hot center and/or disk cutout.
Neither have we seen the putative inner ``refraction'' from the
trailing short- to long-wave branches of the dispersion relation
proposed by \citet{Mark77}; as already noted, every mode we have
found had a strong leading-wave component.

The bar instability of model galaxies, for which theoretical
understanding is steadily improving, persists as an unsolved problem
for real galaxies.  The preference in galaxies for bisymmetric spiral
patterns \citep[see][for a review]{SM22} is interpreted by theorists
as evidence for heavy disks \citep{SC84, ABP87}, but how those same
galaxies can avoid forming a bar still has no satisfactory
explanation, as was highlighted by \citet{SSL} for the case of M33.
Embedding the disk in a sufficiently dense halo that swing
amplification of $m=2$ waves is curtailed does indeed inhibit bar
formation \citep{OP73, ELN}, but would also suppress two armed
spirals, which are the most common patterns in real galaxies
\citep{Davis12, Hart16, YH18}.  The suggestion by \citet{Toom81} that
cutting the feedback loop of the cavity mode would stabilize the disk
seemed promising, but the damping of a mode at an ILR can be
overwhelmed by noise \citep{Sell89, Sell11}.  Furthermore, this paper
has, if anything, exacerbated the puzzle by showing that global
stability is little affected by a dynamically hot center to the disk,
or a deep cutout of responsive mass from the inner disk.

Thus the puzzle of why some galaxies lack bars remains unsolved.
Although real galaxies are clearly affected by non-gravitational
physics, the global stability of galaxy disks must be dominated by
gravitational dynamics.  We therefore believe that the absence of bars
in some galaxies formed in the cosmological simulations (see \S1.2)
has an as yet unidentified dynamical explanation.  One way forward may
then be to try to identify possible stabilizing factors in those
complicated simulations, and to test them one-by-one in more
controlled experiments.

\newpage
\section*{Acknowledgements}
We thank the anonymous referee for providing a thoughtful report based
on a careful read of the paper.  JAS acknowledges the continuing
hospitality and support of Steward Observatory.  RGC acknowledges
support of NSERC grant 2016-05560.

\section*{Data availability}
The data from the simulations reported here can be made available on
request.  The simulation code and analysis software can be downloaded
in one bundle from {\tt http://www.physics.rutgers.edu/galaxy}, and is
documented in the code manual \citep{Sell14}.


\end{document}